\documentclass[fleqn,usenatbib]{mnras}

\usepackage{newtxtext,newtxmath}

\usepackage[T1]{fontenc}
\usepackage{threeparttable}
\usepackage[whole]{bxcjkjatype}
\usepackage{longtable}
\usepackage{url}
\usepackage{comment}

\DeclareRobustCommand{\VAN}[3]{#2}
\let\VANthebibliography\thebibliography
\def\thebibliography{\DeclareRobustCommand{\VAN}[3]{##3}\VANthebibliography}

\usepackage{graphicx}	
\usepackage{amsmath}	

\usepackage{amssymb}	
\usepackage[table,xcdraw]{xcolor}

\newcommand{\scm}{cm$^{-2}$}	
\newcommand{\ccm}{cm$^{-3}$}	

\newcommand{\kms}{km s$^{-1}$}
\newcommand{\HII}{H{\sc ii}\,}
\newcommand{\HI}{H{\sc i}\,}
\newcommand{\co}{$^{12}$CO}
\newcommand{\cothirteen}{$^{13}$CO}
\newcommand{\coeighteen}{C$^{18}$O}

\title[Multi log-normal density structure]
{Multi log-normal density structure in Cygnus-X molecular clouds: A fitting for N-PDF without power-law\thanks{The full version of Table \ref{table:cloud_property} is only available in electronic form at the CDS via anonymous ftp to cdsarc.u-strasbg.fr (130.79.125.5) or via \url{http://cdsweb.u-strasbg.fr/cgi-bin/qcat?J/MNRAS/}.}} 

\author[T. Murase et al.]{
Takeru Murase,$^{1}$\thanks{E-mail:\url{t.murase.univ@gmail.com}(TM)}
Toshihiro Handa,$^{1,2}$\thanks{E-mail:\url{handa@sci.kagoshima-u.ac.jp}(TH)}
Ren Matsusaka,$^{1}$
Yoshito Shimajiri,$^{3,4,5}$
Masato I.N. Kobayashi,$^{3}$\newauthor
Mikito Kohno,$^{6,7}$
Junya Nishi,$^{1}$
Norimi Takeba$^{1}$ and
Yosuke Shibata$^{1}$ \\
$^{1}$Department of Physics and Astronomy, Graduate School of Science and Engineering, Kagoshima University, 1-21-35 K\^{o}rimoto,\\ Kagoshima, Kagoshima 890-0065, Japan\\
$^{2}$Amanogawa Galaxy Astronomy Research centre, Kagoshima University, 1-21-35 K\^{o}rimoto, Kagoshima, Kagoshima 890-0065, Japan\\
$^{3}$National Astronomical Observatory of Japan, Osawa 2-21-1, Mitaka, Tokyo 181-8588, Japan\\
$^{4}$ Laboratoire d'Astrophysique (AIM), CEA/DRF, CNRS, Universit\'{e} Paris-Saclay, Universit\'{e} Paris Diderot, Sorbonne Paris Cit\'{e}, 91191\\ Gif-sur-Yvette, France\\
$^{5}$ Kyushu Kyoritsu University, Jiyugaoka 1-8, Yahatanishi-ku,Kitakyushu, Fukuoka, 807-8585, Japan\\
$^{6}$ Astronomy Section, Nagoya City Science Museum, 2-17-1 Sakae, Naka-ku, Nagoya, Aichi, 460-0008, Japan\\
$^{7}$ Department of Physics, Graduate School of Science, Nagoya University, Furo-cho, Chikusa-ku, Nagoya, Aichi 464-8602, Japan
}

\date{Accepted XXX. Received YYY; in original form ZZZ}

\pubyear{2023}

\begin{document}
\label{firstpage}
\pagerange{\pageref{firstpage}--\pageref{lastpage}}
\maketitle

\begin{abstract}
We studied the H$_2$ column density probability distribution function (N-PDF) based on molecular emission lines using the Nobeyama 45-m Cygnus X CO survey data. Using the DENDROGRAM and SCIMES algorithms, we identified 124 molecular clouds in the \cothirteen\ data. From these identified molecular clouds, an N-PDF was constructed for 11 molecular clouds with an extent of more than 0.4 deg$^2$. From the fitting of the N-PDF, we found that the N-PDF could be well-fitted with one or two log-normal distributions. These fitting results provided an alternative density structure for molecular clouds from a conventional picture. 
We investigated the column density, dense molecular cloud cores, and radio continuum source distributions in each cloud and found that the N-PDF shape was less correlated with the star-forming activity over a whole cloud. Furthermore, we found that the log-normal N-PDF parameters obtained from the fitting showed two impressive features. First, the log-normal distribution at the low-density part had the same mean column density ($\sim$ 10$^{21.5}$ \scm) for almost all the molecular clouds. Second, the width of the log-normal distribution tended to decrease with an increasing mean density of the structures. These correlations suggest that the shape of the N-PDF reflects the relationship between the density and turbulent structure of the whole molecular cloud but is less affected by star-forming activities.
\end{abstract}




\section{Introduction}
\label{sec:intro}
Molecular clouds are an important evolutionary stage in star formation scenarios; they contain dense regions directly linked to star formation and diffuse regions \citep[e.g.][]{Larson1994,Larson1995,Stutzki1998,Elmegreen2000}. Although gravitational collapse is an essential mechanism, it is still open to evolving from an initial density fluctuation to a dense core; the proposed initial processes are turbulence, self-gravity, and massive star feedback \citep{McKee2007}.

The volume density probability distribution function ($\rho$-PDF) has been used as an effective tool for investigating the density structure of molecular clouds affected by various physical processes \citep[e.g.][]{Vazquez-Semadeni1994,Vazquez-Semadeni2001,Ostriker2001,Ballesteros-Paredes2011,Federrath2013,Federrath2016,Jaupart2020,Khullar2021}. The shapes of the $\rho$-PDF are determined by the dominant physical processes governing the density structure of molecular clouds. Theoretical studies of supersonic turbulence suggest that the probability density function of the gas density, or the $\rho$-PDF, of turbulence-dominated molecular clouds exhibits a log-normal distribution \citep[e.g.][]{Vazquez-Semadeni1994,Passot1998,Ostriker2001,Federrath2008,Burkhart2015,Nolan2015}. This feature is mainly attributed to the application of the central limit theorem to the hierarchical density fields formed by multiplicative processes \citep[see][]{Vazquez-Semadeni1994,Passot1998,McKee2007}. When the dominant process is asymmetric in density evolution, such as gravitational collapse, its $\rho$-PDF has an excess tail. This is a traditional interpretation of the power-law tail of the $\rho$-PDF at the high-density end \citep[e.g.][]{klessen2000,Ballesteros-Paredes2011,Kritsuk2011,Collins2012,Federrath2013,Burkhart2015}. However, it is difficult to obtain the $\rho$-PDF of a molecular cloud based on observations owing to the line-of-sight projection effect. Therefore, observational studies have used mainly the PDF of column density, or N-PDF.

Three tracers have been used in observational studies: dust extinction in the near-infrared \citep[e.g.][]{Goodman2009,Kainulainen2009}, dust emission in the mid/far-infrared \citep[e.g.][]{Goodman2009,Schneider2012,Schneider2013,Schneider2015a,Schneider2015b,Schneider2015c,Konyves2015,Lombardi2015}, and optically thin molecular line emission \citep[e.g.][]{Goldsmith2008,Goodman2009,Schneider2016,Ma2020,Wang2020,Ma2021,Ma2022}. Previous studies using these three tracers have shown that N-PDFs of many quiescent clouds are log-normal, whereas those of active star-forming clouds consist of two components: a log-normal and a power-law tail in the high-density range \citep[e.g.][]{Goodman2009,Kainulainen2009,Schneider2013,Lewis2022}. 
However, it has been reported that approximately 70\% of the molecular clouds in the second and third quadrants of the Milky Way mid-plane are log-normal N-PDF only \citep{Ma2021,Ma2022}. This means that most molecular clouds are dominated by turbulence. In addition, they pointed out that power-law N-PDFs are not necessarily associated with star formation activity \citep{Ma2021}.

Far-infrared observations and simulations with high resolution have a third component: another power-law tail at the highest density end \citep{Kritsuk2011,Schneider2015a,Pokhrel2016,Jaupart2020,Khullar2021}. Simulations of isothermal gravitational turbulence performed by \cite{Khullar2021} suggested that the resultant PDF has a log-normal and two power-law tails. The first power-law tail in the intermediate-density range with a steeper slope is due to the onset of self-gravity, and the second tail in the denser range is due to the onset of rotational support. However, it is unclear whether the three components of an observational N-PDF correspond to the components owing to these physical processes \citep[e.g.][]{Tremblin2014,Schneider2015a,Pokhrel2016}. 

We highlight that the scale size of structures attributed to the power-law N-PDF is different between observations and numerical simulation studies in order of magnitude (see Section \ref{sec:absence_pl} in detail). Therefore, the ``so-called'' power-law components of the observed N-PDF may not be of power-law origin. Only a few observational studies on N-PDF have used multi-log-normal fitting. \cite{Tremblin2014} have successfully performed a two log-normal plus one power-law tail fitting for N-PDFs of four molecular clouds associated with \HII regions using \textit{Herschel} dust continuum data \citep{Motte2010,Motte2012}. They interpreted the components of an N-PDF as follows: the log-normal in the lowest-density range comes from the turbulence of a molecular gas, the log-normal in the intermediate-density range comes from the compressed area induced by the expansion of the \HII regions into the first component, and the power-law tail comes from the gravitationally collapsing cores \citep{Tremblin2014}. However, no investigation of N-PDF using multi-log-normal functions has been performed on molecular clouds in various environments. Are \HII regions required to produce the second log-normal of the N-PDF for a molecular cloud? Are power-law tails required for an N-PDF, even when we employ multi-log-normal components?
To investigate this, we performed an N-PDF study of molecular clouds based on a fit using a multi-log-normal distribution.

Observations of dust continuum emission have been used in N-PDF studies. However, since a large number of molecular clouds are located near the galactic plane, it is difficult to study the N-PDF of each cloud by continuum emission because of the large contamination by foreground and background emission.
We, therefore, made N-PDFs from the molecular line data instead of the dust continuum. It is because the molecular line emission provides kinematic information on molecular clouds, allowing the separation of overlapping molecular clouds by projection effects. 

The remainder of this paper is organised as follows. First, we describe the decomposition of the molecular line emission into individual clouds and the procedure for creating an N-PDF for each molecular cloud (Section \ref{sec:analysis}). We present the fit results using the multi-log-normal functions in Section \ref{sec:discussion}. Using the fit results, we discuss the relationship between the shape of the N-PDF and the star-forming activity in the interior of a molecular cloud, as well as the parameters that form the N-PDF. In Section \ref{sec:conclusion}, we summarise our results and conclusions.

\section{target and data sets}

\begin{figure}
	\includegraphics[width=\linewidth, trim={0 180 0 0}, clip]{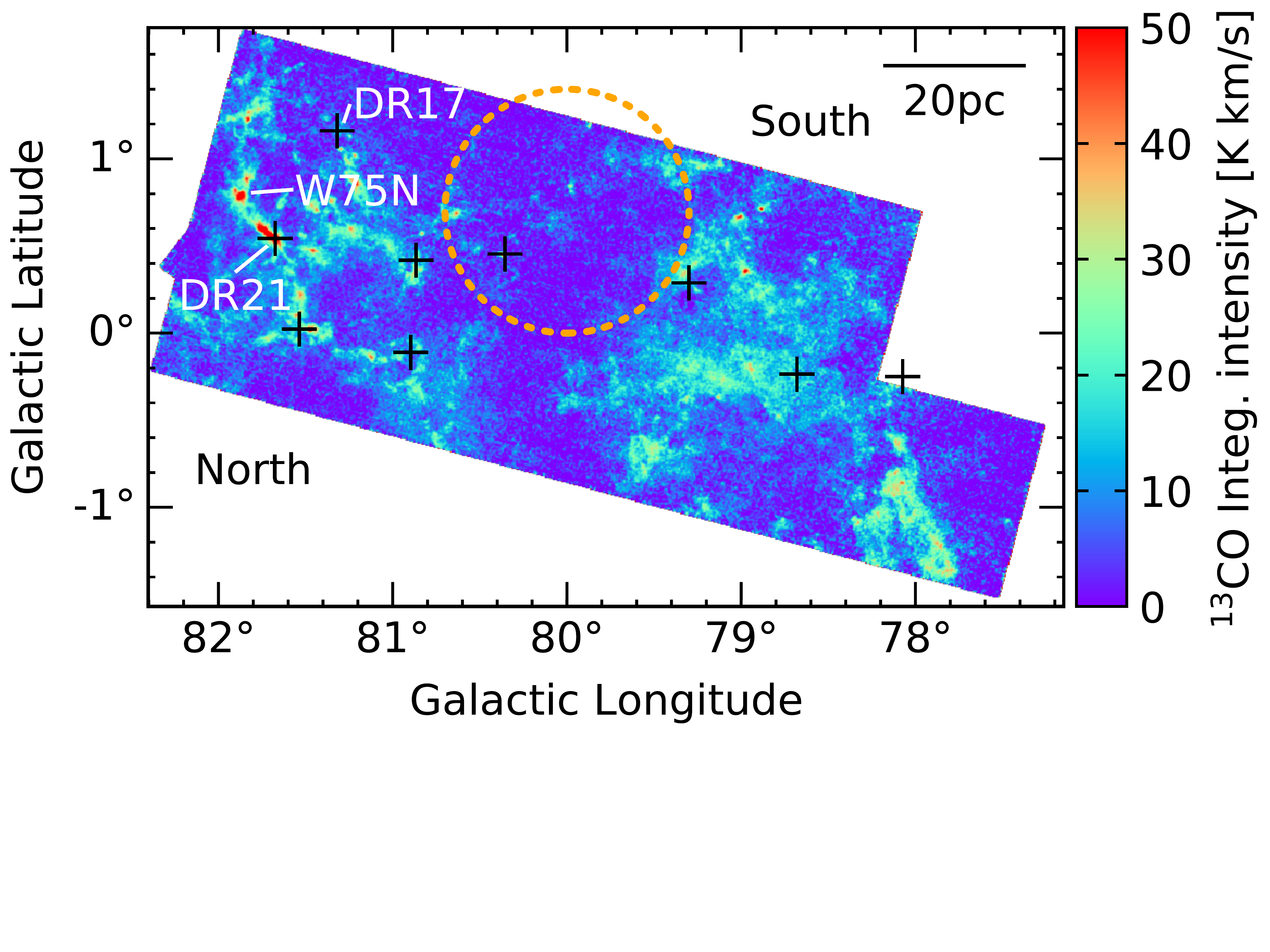}
    \caption{Integrated intensity map in \cothirteen\ ($J$=1--0) obtained by the Nobeyama 45-m Cygnus X survey \citep{Yamagishi2018,Takekoshi2019}. The integrated velocity range is between $-$20 \kms and +20 \kms. The orange dashed line indicates the position of the Cygnus OB2 association.
    Plus marks indicate the positions of 5 GHz radio continuum sources as identified by \citet{Downes1966}.
    }
    \label{fig:integ_map}
\end{figure}

\subsection{Cygnus X complex}
\label{sec:cyg-x}

Cygnus X is a giant molecular cloud (GMC) complex in the Milky Way. The distance to the Cygnus X complex based on annual parallax was established as 1.4 kpc using very long baseline interferometry (VLBI) water maser observations \citep{Rygl2012}. Previous observational studies have conducted large-scale CO surveys \citep[e.g.][]{Schneider2006,Schneider2007,Schneider2016,Yamagishi2018}. The total mass of the molecular gas was estimated to be $3 \times 10^6$ M$_\odot$ based on observations of \cothirteen\ ($J$=2--1) \citep{Schneider2006}. 

The Cygnus X complex contains nine OB associations \citep{Humphreys1978}. The most active star-forming region, which contains 169 OB stars, is the Cygnus OB2 association located in the centre of the Cygnus X complex \citep{Wright2015}. The Cygnus X complex is divided into northern and southern cloud complexes, where the gas masses traced by \cothirteen\ ($J$=2--1) are $2 \times 10^5$ M$_\odot$ and $3 \times 10^5$ M$_\odot$, respectively \citep{Schneider2006}. In this study, the border between the northern and southern regions of Cygnus X was determined at a galactic longitude of 80\degr.2, as \cite{Takekoshi2019}.

Previous studies have reported that these molecular cloud complexes have different star formation properties \citep[e.g.][]{Schneider2006,Schneider2007,Schneider2011,Schneider2016,Yamagishi2018,Takekoshi2019}. \cite{Takekoshi2019} has investigated the statistical properties of dense clumps associated in both regions using \coeighteen\ emission. Consequently, they found that the average radius and the velocity dispersion of the dense clumps were similar in both regions. However, the H$_2$ density was significantly higher in the northern region. This statistical difference suggests a reflection of the star-forming activities in the northern and southern regions \citep[see Figure 4 in][]{Takekoshi2019}.

In the northern part of the Cygnus X complex, many filamentary structures have been identified in dust and molecular gas tracers \citep[e.g.][and see also Figure \ref{fig:integ_map}]{Schneider2006,Motte2010,Hennemann2012,Schneider2016,Schneider2016b,Yamagishi2018}. In addition, there are well-known star-forming regions such as DR17, DR21, and W75N, which are composed of numerous fragmentary structures and massive dense cores that can form high-mass stars. The southern part of the Cygnus X complex exhibits a diffuse structure (Figure \ref{fig:integ_map}). Although the southern part of Cygnus X exhibits weaker star-forming activity than the northern part, the presence of a large amount of molecular gas suggests the possibility of star cluster formation \citep{Schneider2006,Yamagishi2018}. As described above, the Cygnus X complex contains various star-forming environments. Therefore, this target is suitable for investigating the relationships between the density structure of molecular clouds and star-formation activities.

\subsection{Data sets}
\label{sec:dataset}
We used \co\ ($J$=1--0) and \cothirteen\ ($J$=1--0) line data from the Nobeyama 45-m Cygnus X CO Survey\footnote{The data sets are publicly available at \url{https://cygnus45.github.io}} (hereafter, Cygnus X Survey) \citep{Yamagishi2018,Takekoshi2019}. These data were obtained using a multi-beam receiver, FOur-beam REceiver System on the 45-m Telescope \citep[FOREST;][]{minamidani2016} and a Spectral Analysis Machine for the 45-m telescope \citep[SAM45;][]{kuno2011}. Figure \ref{fig:integ_map} shows the integrated intensity map of \cothirteen\ ($J$=1--0). The observation area was 9 deg$^2$ by connecting $1 \times 1$ deg$^2$ mosaic submaps, which covered the Cygnus X north and south regions, as well as the Cygnus OB2 association. The On-The-Fly (OTF) mapping scan parameters were the same as those of the FOREST Unbiased Galactic plane Imaging survey with the Nobeyama 45-m telescope \citep[FUGIN;][]{Umemoto2017}. The effective angular resolution of the data cubes was 46 arcsec, which corresponded to 0.31 pc at 1.4 kpc. The velocity resolution was 0.25 \kms. 
The Cygnus X Survey data were calibrated to the main beam temperature ($T_\mathrm{mb}$) with a main beam efficiency ($\eta_\mathrm{mb}$) of 0.43 for \co\ and 0.45 for \cothirteen, respectively.
The typical noise levels of \co\ and \cothirteen\ data were 0.88 K and 0.36 K in $T_\mathrm{mb}$ scale, respectively. More detailed information on the observations and data reduction are described in \cite{Yamagishi2018}.

To investigate the distributions of dense molecular cores and radio continuum sources, we used the \coeighteen\ core catalogue obtained from the Cygnus X Survey \citep{Takekoshi2019} and the radio continuum source catalogue of the Red Midcourse Space Experiment (MSX) Source (RMS) survey \citep{Urquhart2009}, respectively. \cite{Urquhart2009} performed continuum observations at 4.86 GHz toward the RMS sources with the Very Large Array (VLA). There were 38 radio continuum sources identified within the observation area of the CO data used in this study, which were classified as compact or ultra-compact \HII regions. The mean size of these radio continuum sources was 2 arcsec $\times$ 1 arcsec, which corresponded to 0.014 pc $\times$ 0.007 pc at 1.4 kpc.

\section{Analysis and results}
\label{sec:analysis}
\subsection{Identification of molecular clouds in \cothirteen\ data}
\label{sec:dendrogram_scimes}

\begin{figure*}
	\includegraphics[width=0.9\linewidth, trim={0 0 280 0}, clip]{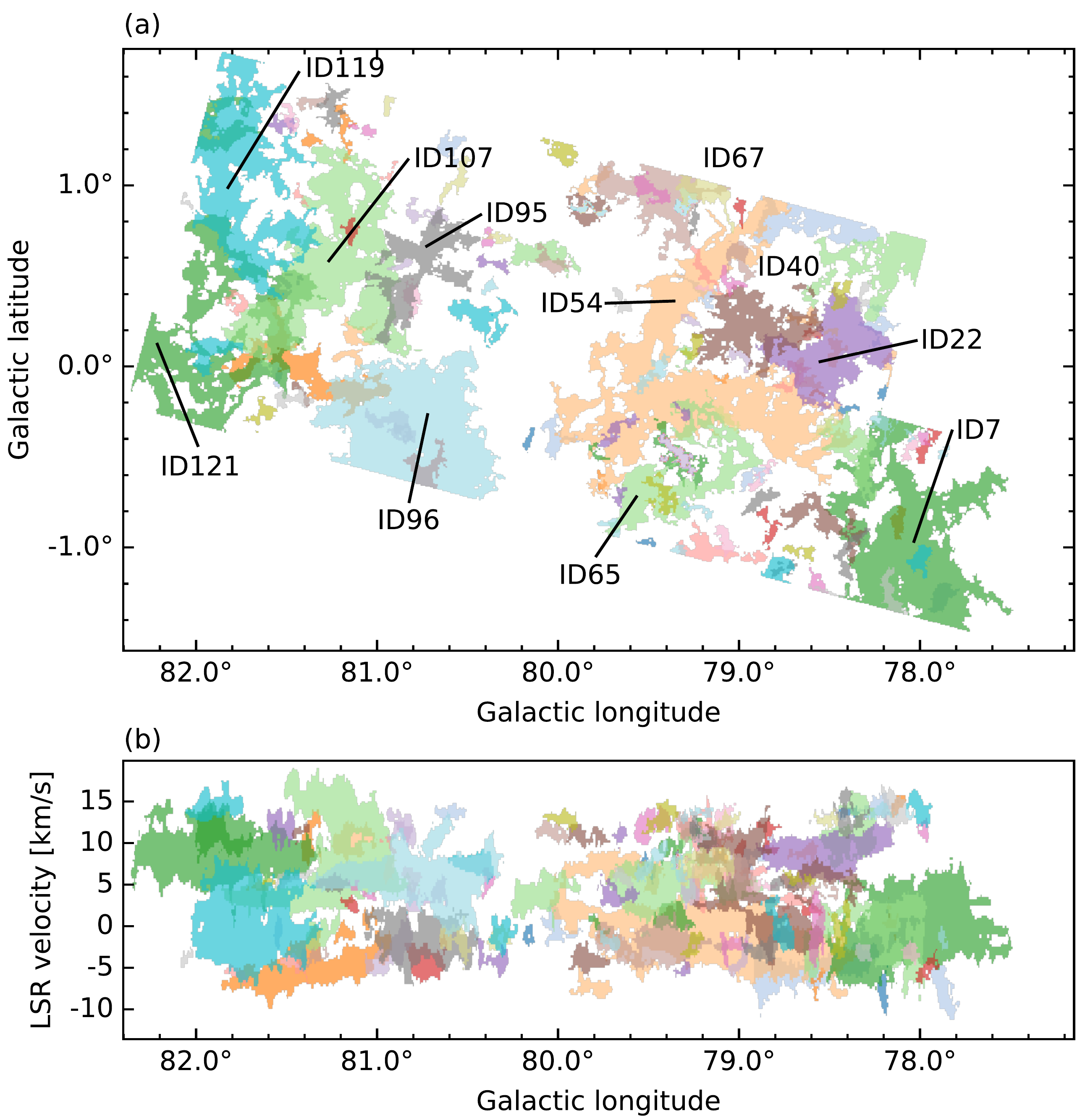}
    \caption{\cothirteen\ emission segmentation performed using DENDROGRAM and SCIMES algorithms. (a) $l$-$b$ space map and (b) $l$-$v$ space map. Each colour shows an individual identified cloud. The segmentation maps use the same colour scheme in panels (a) and (b). ID indicates clouds with an extract projected area over $\sim$ 4000 pixels.}
    \label{fig:identified_structure}
\end{figure*}

Because \cothirteen\ lines generally trace molecular cloud structures better than \co\ lines \cite[e.g.][]{Umemoto2017, Li2018, Ma2021}, we decomposed the molecular line intensity map into individual molecular clouds from \cothirteen\ data cube using DENDROGRAM \citep{Rosolowsky2008} and SCIMES \citep{Colombo2015} algorithms. In this study, we used the \texttt{astrodendro} Python-based package to analyse the dendrograms of the emission line intensity cube. DENDROGRAM is an algorithm that constructs a tree representing a hierarchy of structures comprising a trunk, branches, and leaves \citep[for more information on the DENDROGRAM algorithm, see][]{Rosolowsky2008}. This algorithm requires three input parameters: the \texttt{min\_value}, \texttt{min\_delta}, and \texttt{min\_npix}. Here, \texttt{min\_value} is a threshold to distinguish between line emission and noise, \texttt{min\_delta} represents the threshold for separating two `leaves', and \texttt{min\_npix} is the minimum number of pixels to identify an individual `leaf' structure. We set \texttt{min\_value} = 3$\sigma$ and \texttt{min\_delta} = 2$\sigma$, where $\sigma$ is the RMS noise level of the \cothirteen\ data. We also set \texttt{min\_npix} = 64, corresponding to $4 \times 4 \times 4$ pixels along each of the position-position-velocity (PPV) space.

Next, we applied SCIMES using the output of the DENDROGRAM algorithm. SCIMES is an algorithm that identifies clustered structures (see \citet{Colombo2015} in detail). This analysis used the `volume' as a clustering criterion for the SCIMES algorithm. The outputs from SCIME are PPV mask cubes of the identified clustered structures and a catalogue containing the physical information of each clustered structure. In this paper, we call the clustered structure identified by SCIMES an `individual cloud'.

Based on DENDROGRAM and SCIMES analyses, 131 clustered structures were identified. However, seven compact clusters (extracted area < 100 pixels, corresponding to 1.5 pc$^2$) located at the edge of the map were excluded as noise structures because they were affected by bad channels in the map-making process. Hence, 124 clustered structures were identified as molecular clouds. We present the resulting decomposition of the \cothirteen\ data cube in Figure \ref{fig:identified_structure}. The identified cloud properties are summarised in Table \ref{table:cloud_property}.

\subsection{Deriving column density maps and N-PDFs}
\label{sec:columndesity}

\begin{figure*}
\includegraphics[width=0.45\textwidth, trim={0 20 0 0}, clip]{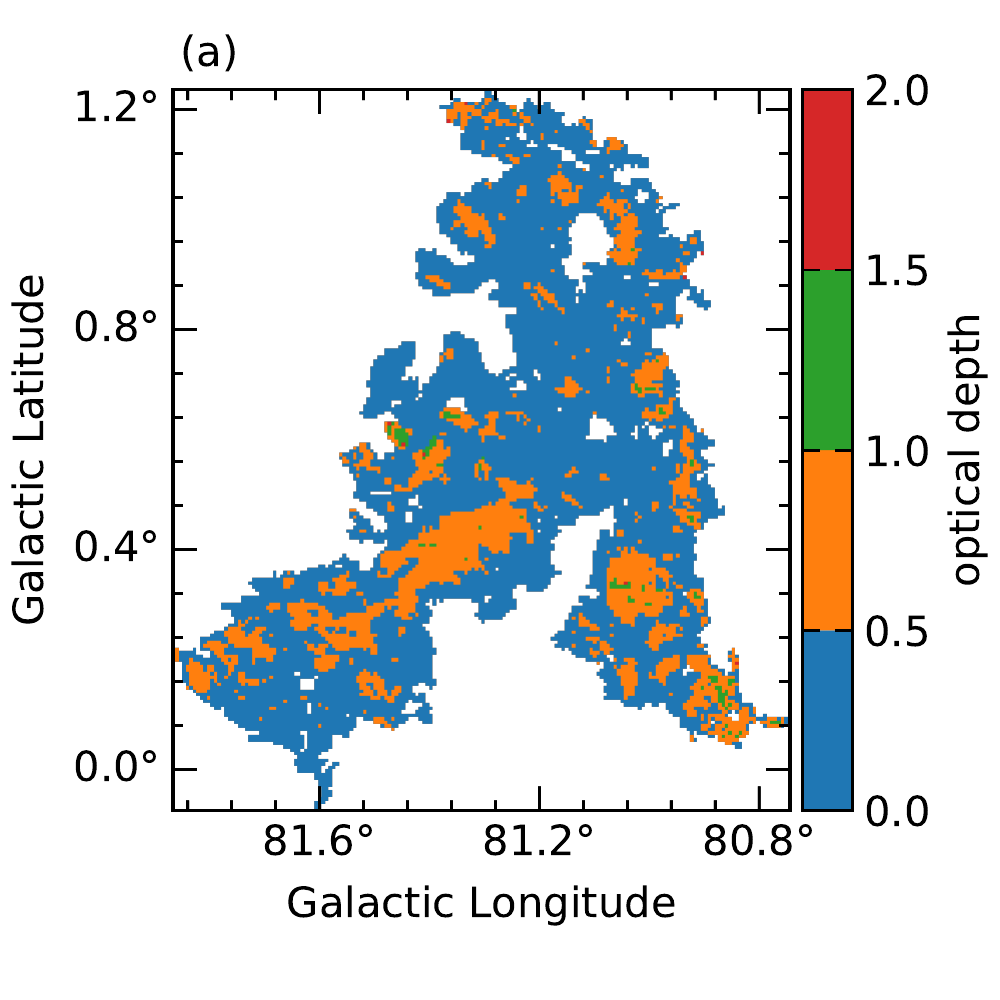}
\includegraphics[width=0.45\textwidth, trim={0 20 0 0}, clip]{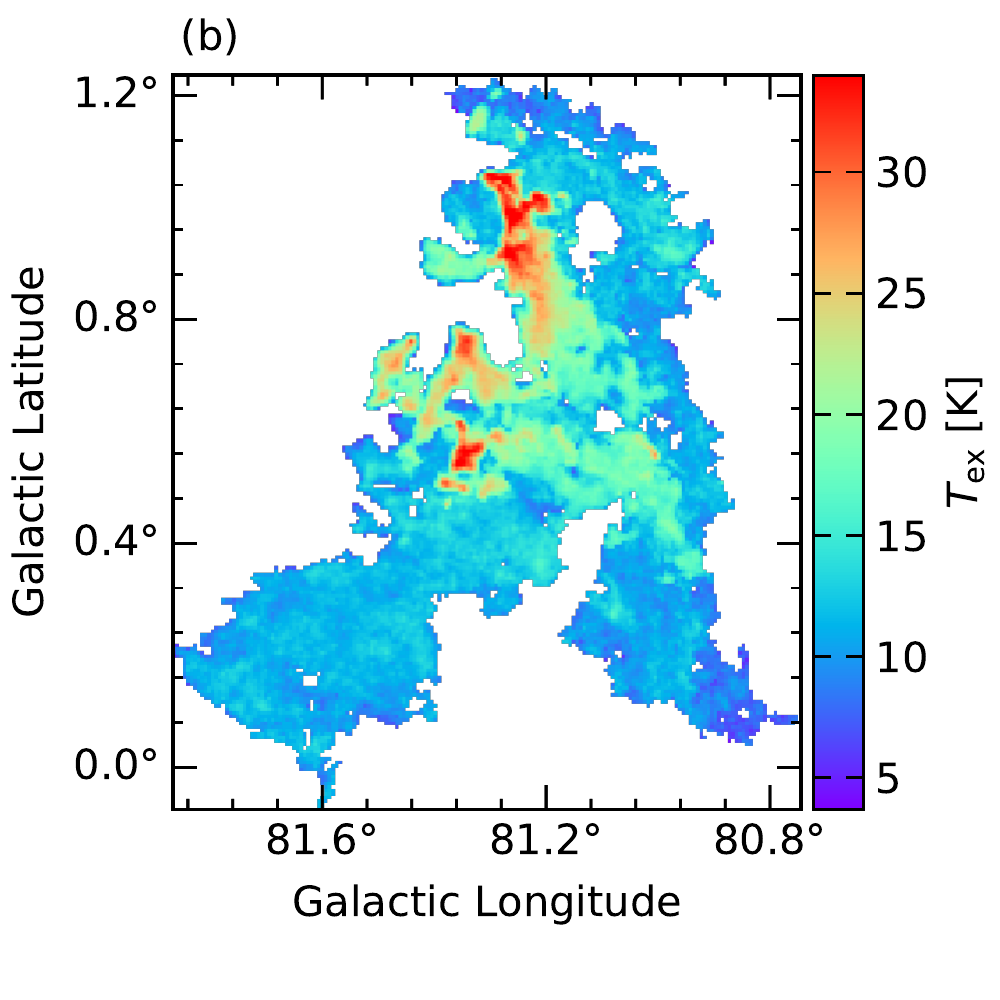}
\includegraphics[width=0.45\textwidth, trim={0 20 0 0}, clip]{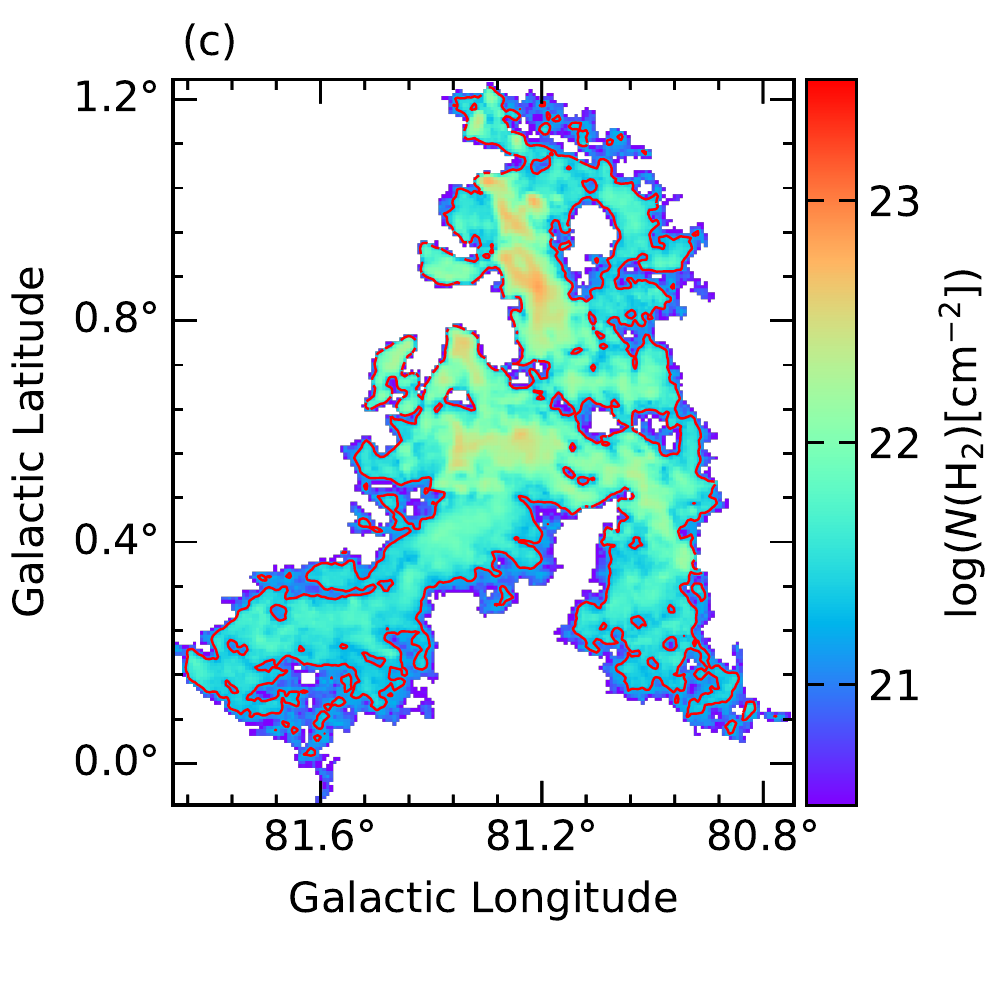}
\includegraphics[width=0.45\textwidth, trim={5 10 10 5}, clip]{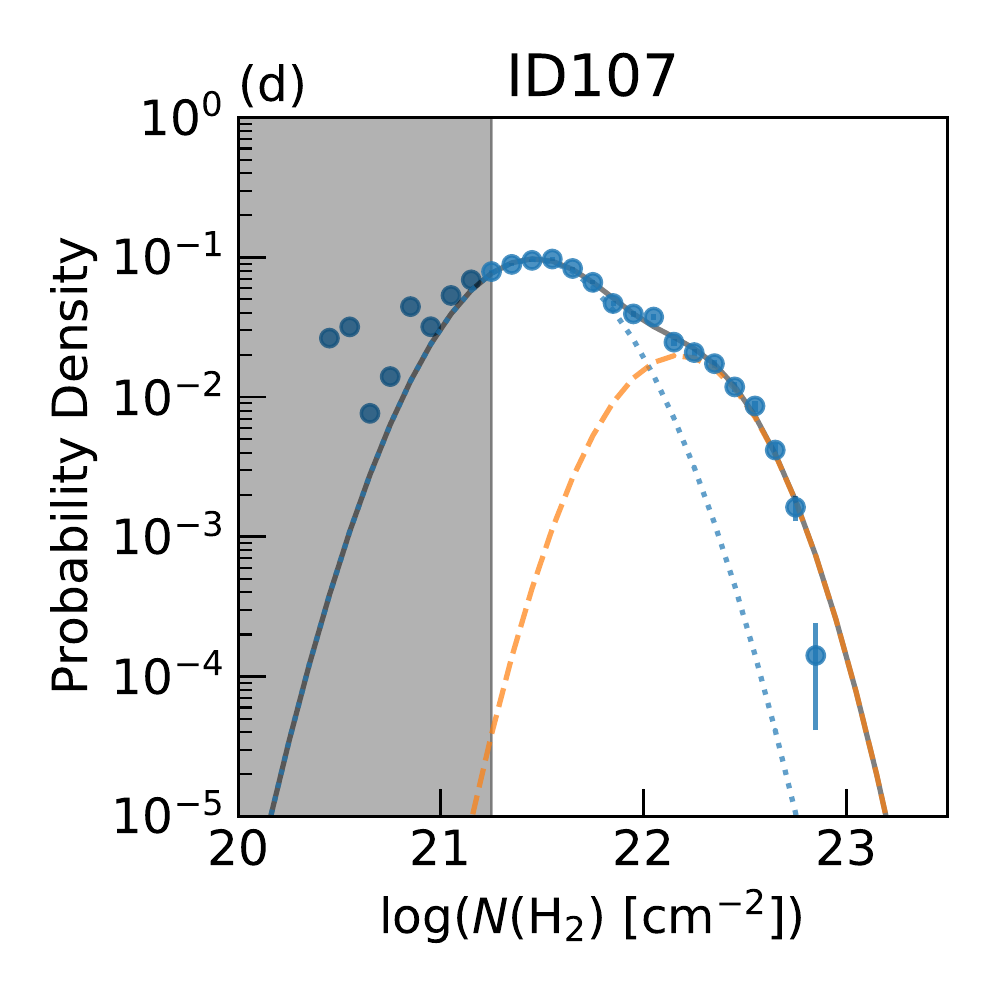}
    \caption{The optical depth, excitation temperature, column density maps of ID107 cloud and its N-PDF. (a) Optical depth map. The colours indicate the optical depth in increments of 0.5. (b) Excitation temperature map obtained from \co\ ($J$=1--0) line. (c) H$_2$ column density map (see column density maps of other selected clouds in Figures \ref{fig:ex_columndensity_1} and \ref{fig:other_columndensity_maps}). The red contour indicates the completeness limit of H$_2$ column density (log($N$(H$_2$)) = 21.25) for the ID107 cloud. (d) N-PDF is derived from the H$_2$ column density map of the panel (c). The grey-shaded areas correspond to column densities out of the completeness limit.}
    \label{fig:ex_columndensity}
\end{figure*}

\begin{table}
    \caption{Cloud parameters identified using SCIMES.}
    \label{table:cloud_property}
    \begin{threeparttable}
    \begin{tabular}{cccccc}
    \hline
        ID   &
        \begin{tabular}[c]{@{}c@{}}$l$\\ {[} deg. {]}\end{tabular} & \begin{tabular}[c]{@{}c@{}}$b$\\ {[} deg. {]}\end{tabular} & \begin{tabular}[c]{@{}c@{}}$v_\mathrm{LSR}$\\ {[} \kms {]}\end{tabular} & \begin{tabular}[c]{@{}c@{}}$\sigma_v$\tnote{a}\\ {[} \kms {]}\end{tabular} &
        \begin{tabular}[c]{@{}c@{}}$S_\mathrm{exact}$\\ {[} pix.$^2$ {]}\end{tabular} \\
    \hline \hline
        7   & 78.063 & $-$1.009 & $-$0.15 & 2.34  & 14202 \\
        22  & 78.479 & 0.105  & 8.63  & 1.30  & 5915  \\
        40  & 78.916 & 0.223  & 4.06  & 4.58  & 4492  \\
        54  & 79.235 & $-$0.036 & $-$0.16 & 2.92  & 21290 \\
        65  & 79.400 & $-$0.636 & 4.68  & 1.48  & 5631  \\
        67  & 79.457 & 0.938  & $-$2.32 & 1.08  & 3937  \\
        95  & 80.776 & 0.511  & $-$1.97 & 1.40  & 4231  \\
        96  & 80.801 & $-$0.337 & 5.48  & 1.74  & 13787 \\
        107 & 81.243 & 0.621  & 7.89  & 4.90  & 14139 \\
        119 & 81.734 & 0.964  & 0.11  & 3.04  & 11051 \\
        121 & 81.839 & 0.255  & 8.65  & 1.66  & 12211 \\      
    \hline
\end{tabular}
\begin{tablenotes}
\item Notes: Only the relatively extended molecular clouds obtained N-PDFs are provided here; the full table is available in electronic format. 
\item[a] The intensity-weighted velocity full width at half maximum.
\end{tablenotes}
\end{threeparttable}
\end{table}

We estimated the column density of \cothirteen\ ($J$=1--0) based on the assumptions that the molecular clouds are under local thermodynamic equilibrium (LTE) with \co\ ($J$=1--0) and the beam filling factors of both lines are unity. We estimated the excitation temperature $T_\mathrm{ex}$ in K using the following equation:
\begin{equation}
\label{eq:excitation}
    T_\mathrm{ex} = 5.53 
    \Big / 
    \ln{\left(1 + \frac{5.53}{T\mathrm{_{mb}^{peak}}(^{12}{\rm CO}) + 0.819} 
    \right)},
\end{equation}
where $T\mathrm{_{mb}^{peak}}(^{12}{\rm CO})$ is the \co\ peak intensity in K for each pixel, which was derived from 
\begin{equation}
    T_\mathrm{MB}=J_{12}(T_\mathrm{ex})-J_{12}(T_\mathrm{CMB}),
\end{equation}
where $J_{12}(T)$ is the brightness temperature in the Rayleigh-Jeans approximation of radiation from a black body with temperature $T$ at the \co\ ($J$=1--0) frequency, and $T_\mathrm{CMB}=2.7$ K is the temperature of the cosmic background radiation. We assumed $T_\mathrm{ex}$ was uniform along the line-of-sight. At this $T_\mathrm{ex}$ in K, we estimated the optical depth of \cothirteen\ at velocity $v$, $\tau_v(\rm {^{13}CO})$, using the following equation:
\begin{equation}
\label{eq:tau}
    \tau_v({\rm ^{13}CO}) = 
    -\ln{ \left[ 1 - \frac{T_\mathrm{mb}({\rm ^{13}CO}) / 5.29 }{1 / \{{e^{5.29 / T_\mathrm{ex}}-1\}} - 0.164} \right] },
\end{equation}
which was derived from 
\begin{equation}
    T_\mathrm{13}=
    \left(J_{13}(T_\mathrm{ex})-J_{13}(T_\mathrm{CMB})\right)(1-e^{-\tau(^{13}\mathrm{CO})}),
\end{equation}
where $J_{13}(T)$ is the brightness temperature at \cothirteen\ ($J$=1--0) frequency. From the velocity-integrated optical depth to the column density $N({\rm^{13}CO})$ in \scm, we used the following equation:
\begin{equation}
\label{eq:column}
    N({\rm^{13}CO}) = 2.42 \times 10^{14}
    \frac{T_\mathrm{ex}+0.88}{1-e^{-5.29/T_\mathrm{ex}}}
    \int \tau_v({\rm ^{13}CO}) dv,
\end{equation}
where $v$ is in \kms. This was derived from Einstein's coefficient model with CMB and LTE assumption for all \cothirteen\ excitation levels. Note that we used the \co\ data cube applied to the PPV mask cube obtained from \cothirteen\ because the shapes of the spectra of \co\ and \cothirteen\ are generally similar \citep[e.g.][]{Heyer2015,Umemoto2017,Li2018,Takekoshi2019}. Here, Eq. (\ref{eq:excitation}) assumes that the \co\ emission is optically thick. The estimated $N$(\cothirteen) was converted into the molecular hydrogen column density, ($N$(H$_2$)), assuming the relative abundances of [$^{12}$C/$^{13}$C] = 70 \citep[derived from local ISM in][]{Wilson1994} and [H$_2$/\co] = 1.1 $\times$ 10$^{4}$, derived from the Taurus molecular cloud \citep{Pineda2010}. Consequently, we used [H$_2$/\cothirteen] abundance of 7.7 $\times$ 10$^{5}$ in this study. Note that the resultant N-PDF shifts only along the abscissa if the abundance value is different.

To obtain an accurate N-PDF, we chose 11 clouds with an extracted projected area of over $\sim$ 4000 pixels, i.e., 0.4 deg$^2$. We show the optical depth, excitation temperature, column density map, and N-PDF of one of the 11 clouds in Figure \ref{fig:ex_columndensity}. In Figure \ref{fig:ex_columndensity}-(a), most areas were found to be optically thin ($\tau$(\cothirteen) < 0.5) and some pixels were optically thick ($\tau$(\cothirteen) > 1). 
Although such pixels with large optical depth may refer to clumps and cores, their effect on the N-PDF was considered to be small due to the small number of pixels in the cloud.
The median optical depth of each molecular cloud ranged from 0.3 to 0.5 (Column 2 in Table \ref{tab:N-PDF_params}). Figure \ref{fig:ex_columndensity}-(b) shows the excitation temperature map. The range of excitation temperature was between 4 and 33 K. We can find that regions with relatively high temperatures ($T_\mathrm{ex}$ > 25 K) are concentrated, and each of them is expected to be an active star-forming region \citep[e.g.][]{Murase2022}. We summarised the physical parameters of the 11 clouds in Table \ref{tab:N-PDF_params}.

\cite{Schneider2015b} reported that the size of bins and angular resolution did not affect the N-PDF parameters (e.g. mean column density and width). In this study, we used a bin size of 0.1 dex. To obtain the correct N-PDF, it is necessary to consider the completeness limit of the column-density map. \cite{Alves2017} have recommended that the completeness limit is determined by the lowest closed contour of the column density map \citep[see also][]{Lombardi2015,Kortgen2019}. Therefore, we manually determined the completeness limit of the column density map for each cloud. As a result, we found the contours at log($N$(H$_2$)) = 21.25, which are indicated by the red contours in Figure \ref{fig:ex_columndensity}-(c), were almost closed. The clouds near the edge of the observation area had almost closed contours at $\log(N(\mathrm{H}_2))=21.5$. It should be noted that the completeness limit has to be set to a higher column density for a column density map that includes the edge of the observation area. Even though the contour at the completeness limit was not completely closed, the effect on the PDF shape was considered small because the number of pixels contained in the column density bin around the completeness limit was large (see Figure \ref{fig:N-PDFs}). Five out of the eleven clouds were detached from the edge of the observation area and had closed boundaries (Column 12 in Table \ref{tab:N-PDF_params}). It supports us in making an appropriate N-PDF fitting, as shown in the next subsection.

On the other hand, it should be noted that molecular clouds that contain the edge of the observation area do not have correct N-PDFs due to missing samples \citep[see also][]{Alves2017}. This effect is seen in the N-PDFs of ID7, ID96, and ID119 clouds in the low-density range. Furthermore, we found a cutoff feature in the N-PDF of ID119 cloud at log($N$(H$_2$)) > 23. The ID119 cloud contains an active massive star-forming region DR21. In such a region, local heating or photodissociation by UV radiation from the star-forming activity can have a significant influence on the CO abundance \citep{Ripple2013}. For these reasons, the ID119 cloud is not used in the following discussion.

At low temperature ($T$ < 20 K) and high volume density ($n$(H$_2$) > 10$^{4}$ \ccm) regions, such as the infrared dark clouds (IRDCs), the abundance of CO molecules in the gas phase is affected by adsorption effects on the dust grains \citep[e.g.][]{Willacy1998,Tafalla2002}. However, \cite{Hernandez2011a} reported that the ratio of the mass surface density obtained from \cothirteen\ ($J$=1--0) and dust extinction observations at IRDC is approximately constant.

\subsection{Fitting the N-PDFs with multi log-normal distributions}
\label{sec:fit_PDF}
\begin{figure*}
\includegraphics[width=0.33\textwidth, trim={0 15 0 10}, clip]{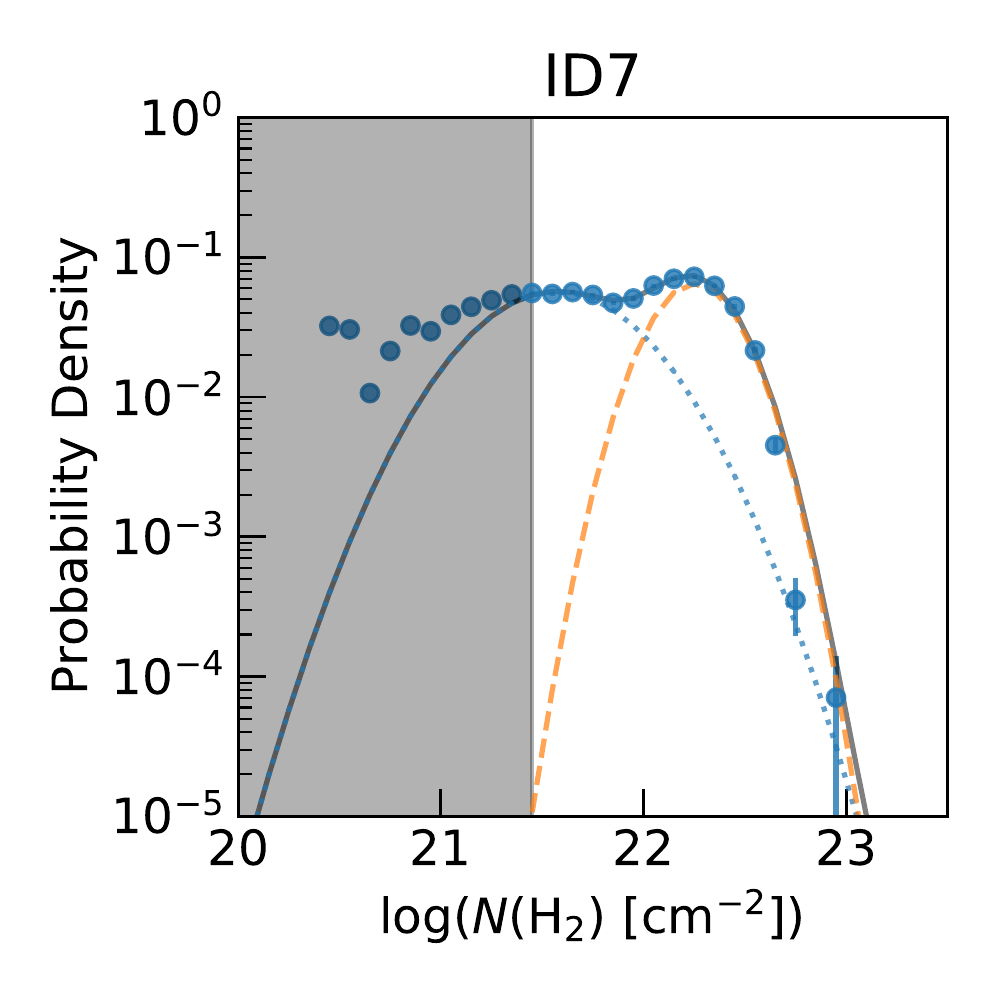}
\includegraphics[width=0.33\textwidth, trim={0 15 0 10}, clip]{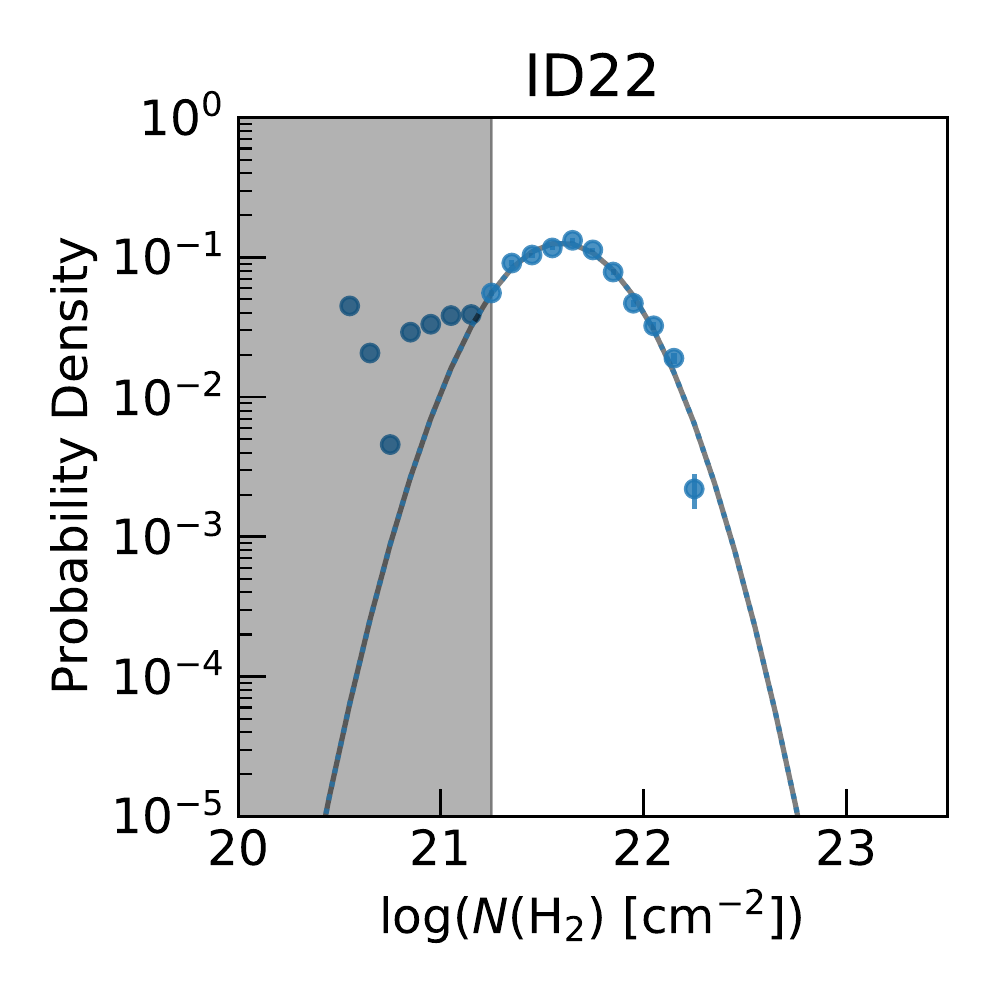}
\includegraphics[width=0.33\textwidth, trim={0 15 0 10}, clip]{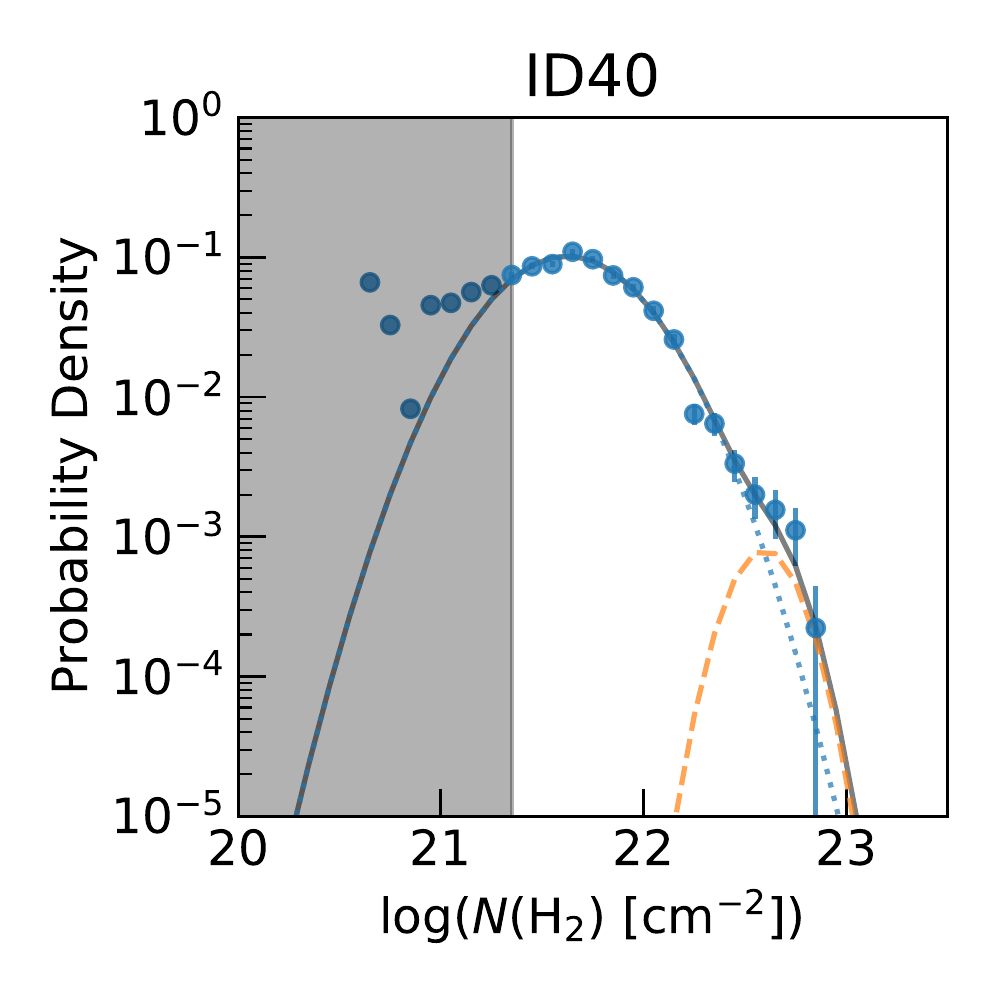}
\includegraphics[width=0.33\textwidth, trim={0 15 0 10}, clip]{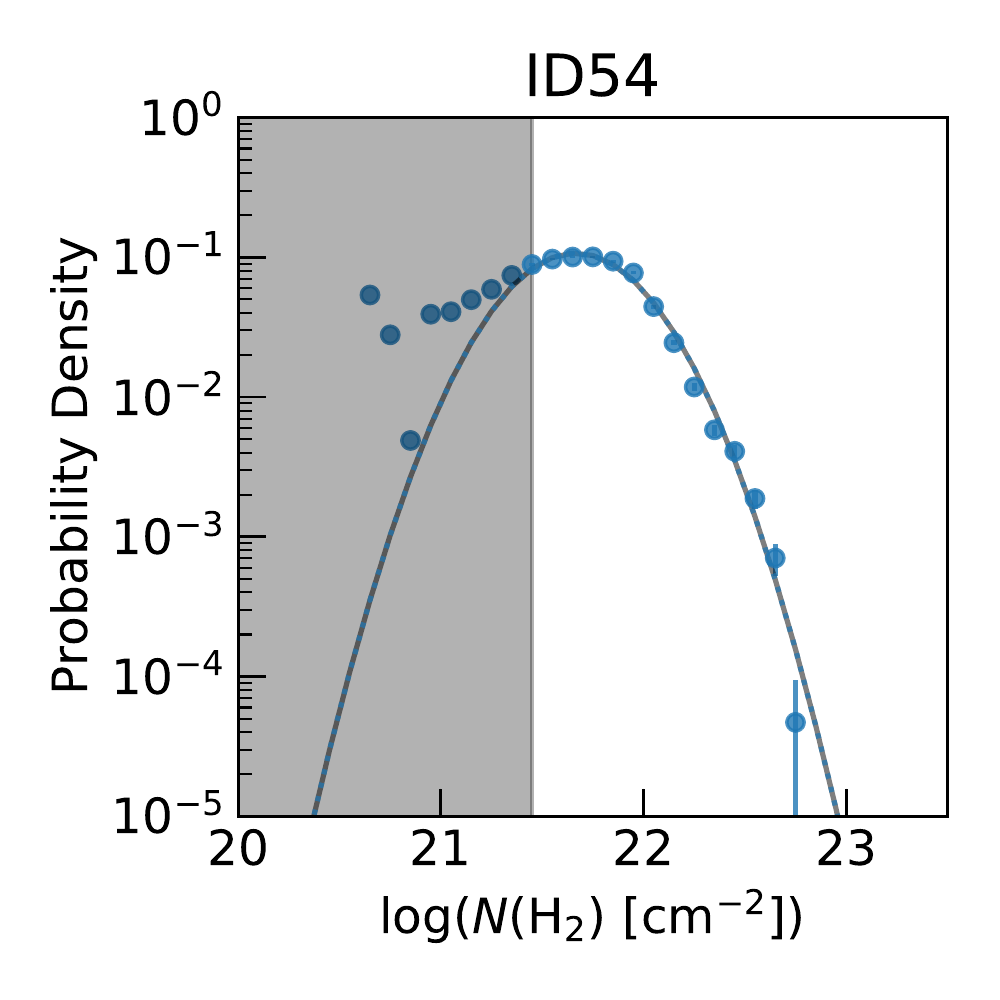}
\includegraphics[width=0.33\textwidth, trim={0 15 0 10}, clip]{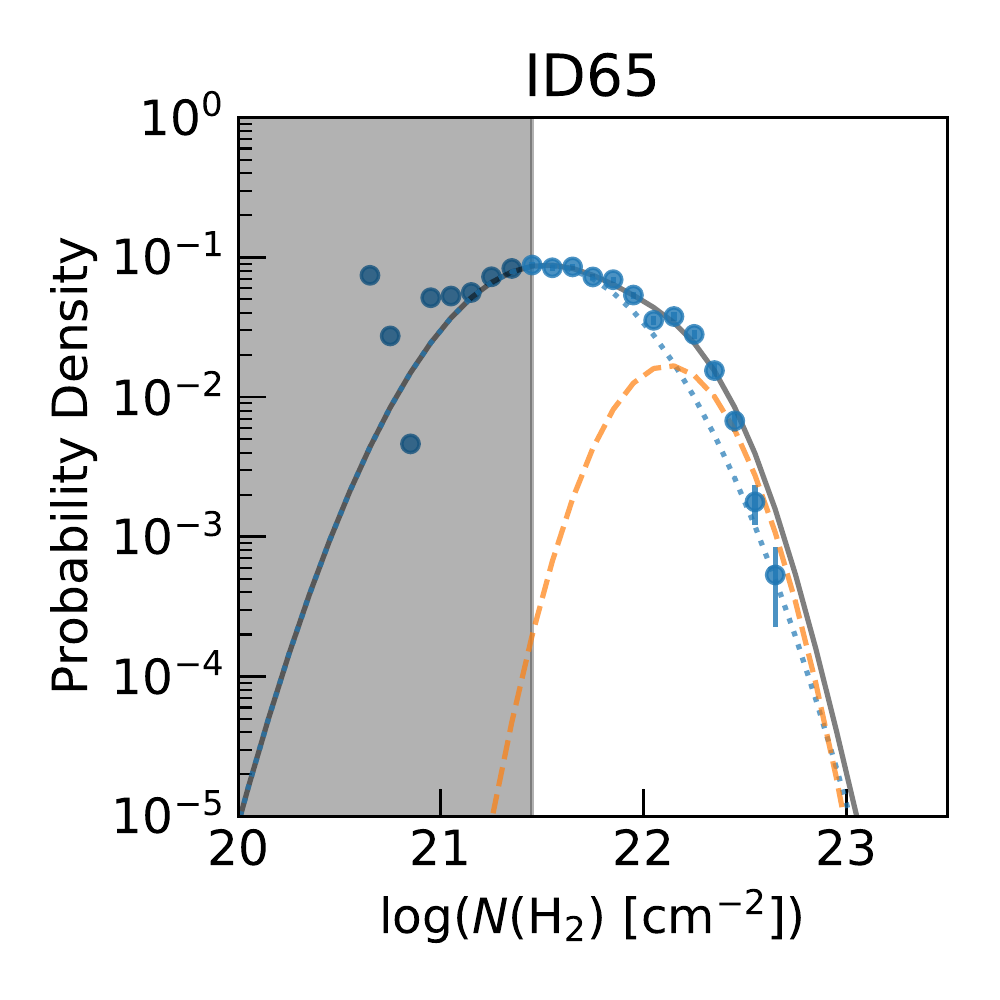}
\includegraphics[width=0.33\textwidth, trim={0 15 0 10}, clip]{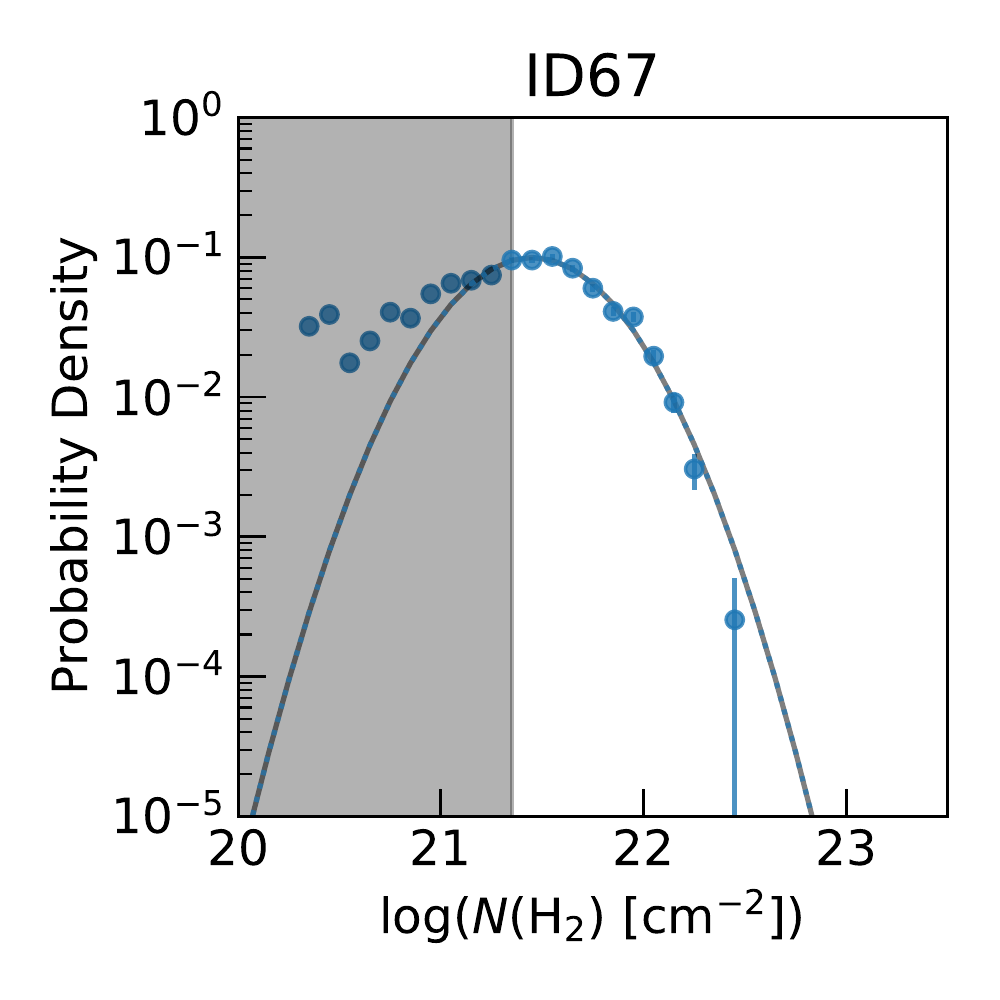}
\includegraphics[width=0.33\textwidth, trim={0 15 0 10}, clip]{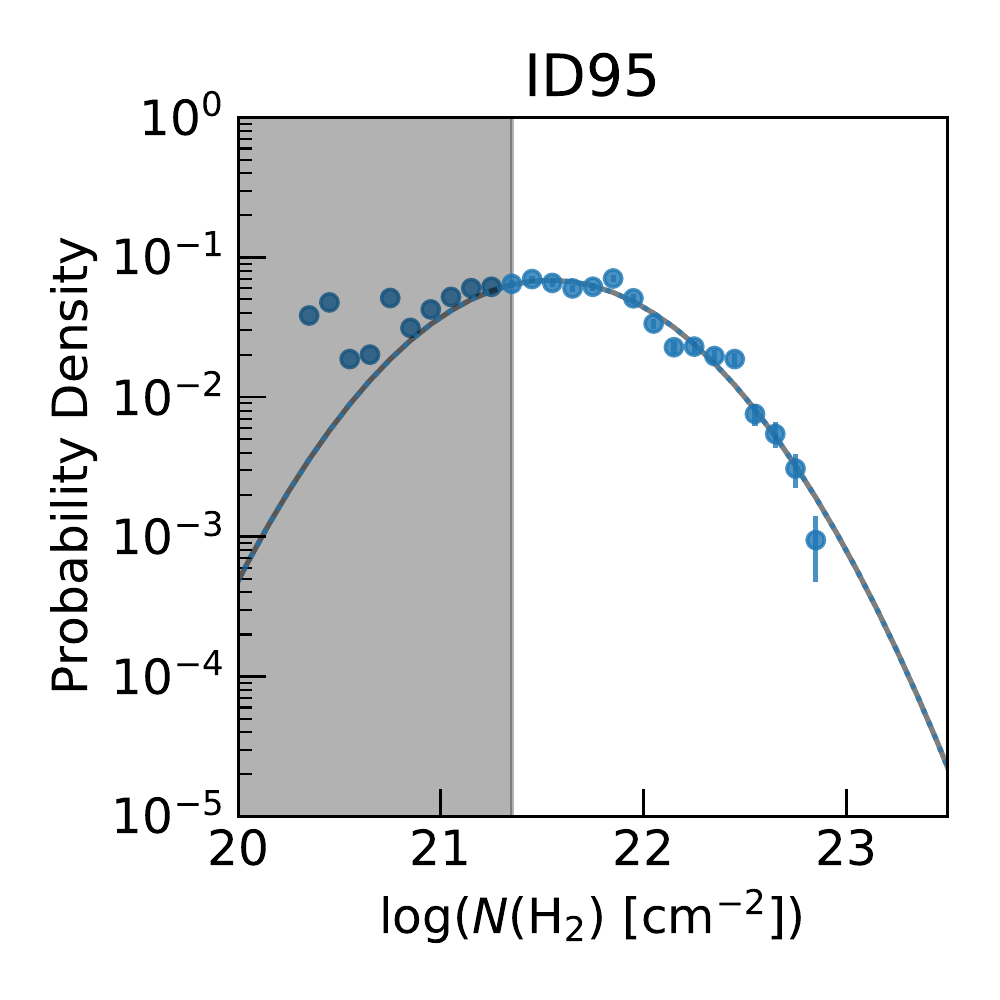}
\includegraphics[width=0.33\textwidth, trim={0 15 0 10}, clip]{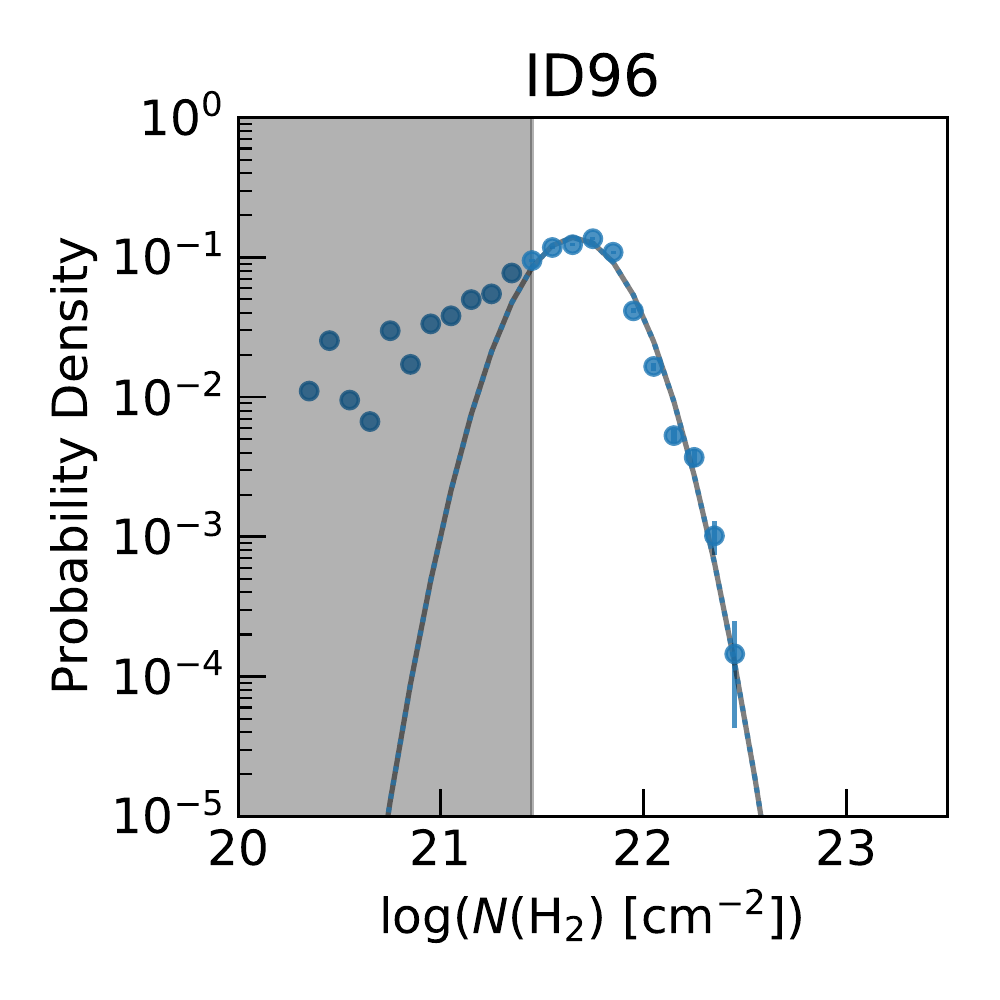}
\includegraphics[width=0.33\textwidth, trim={0 15 0 10}, clip]{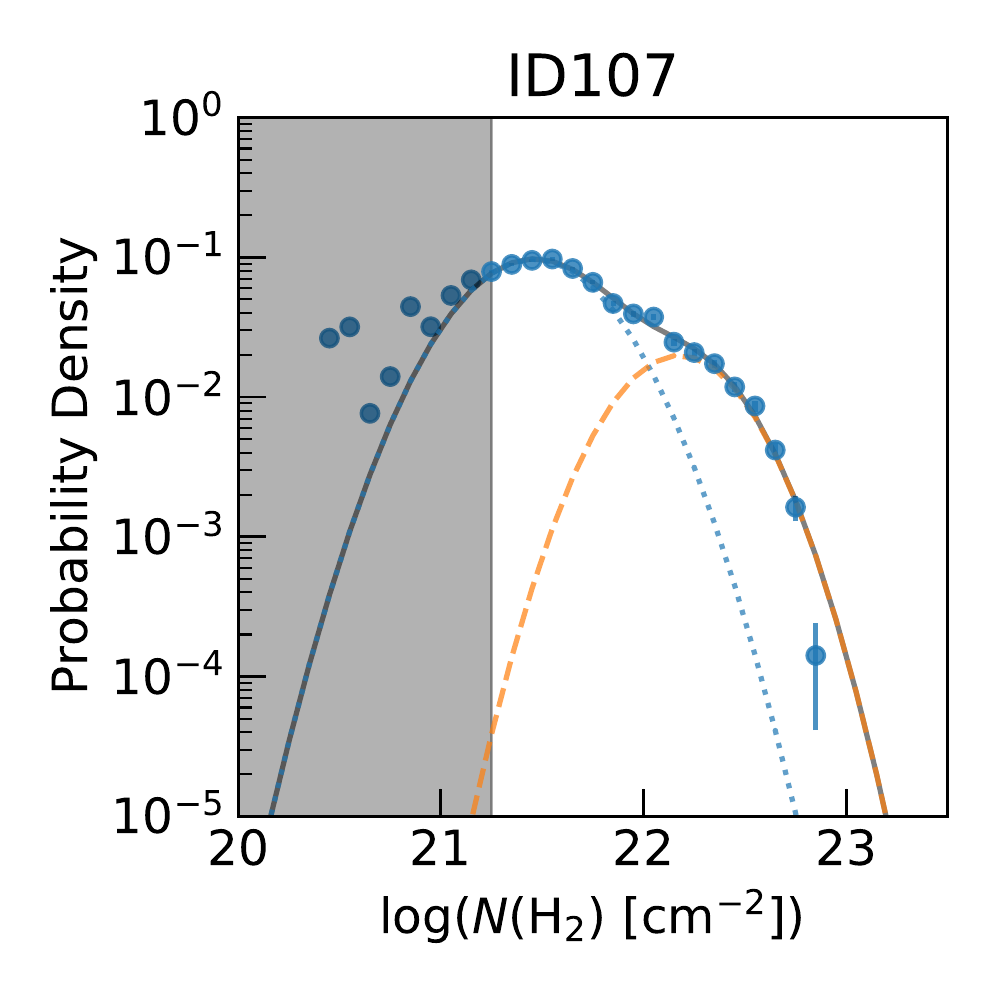}
\includegraphics[width=0.33\textwidth, trim={0 15 0 10}, clip]{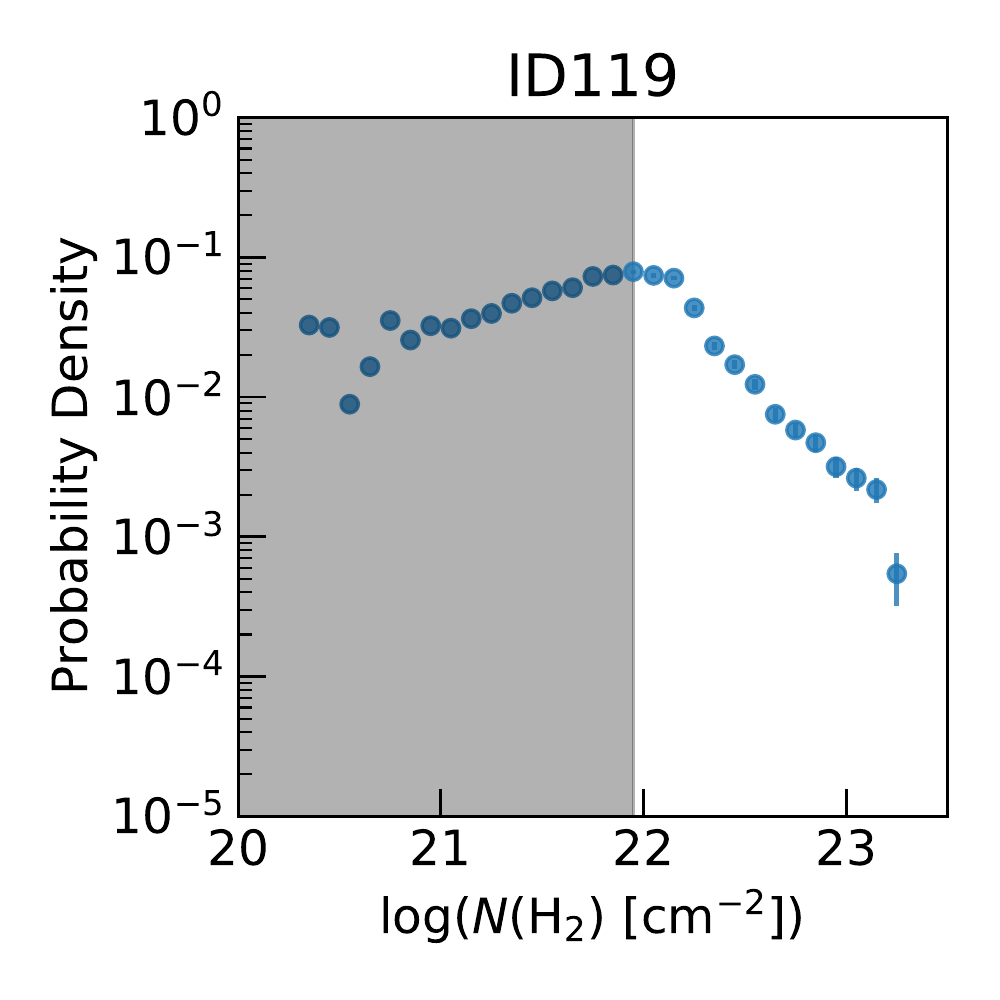}
\includegraphics[width=0.33\textwidth, trim={0 15 0 10}, clip]{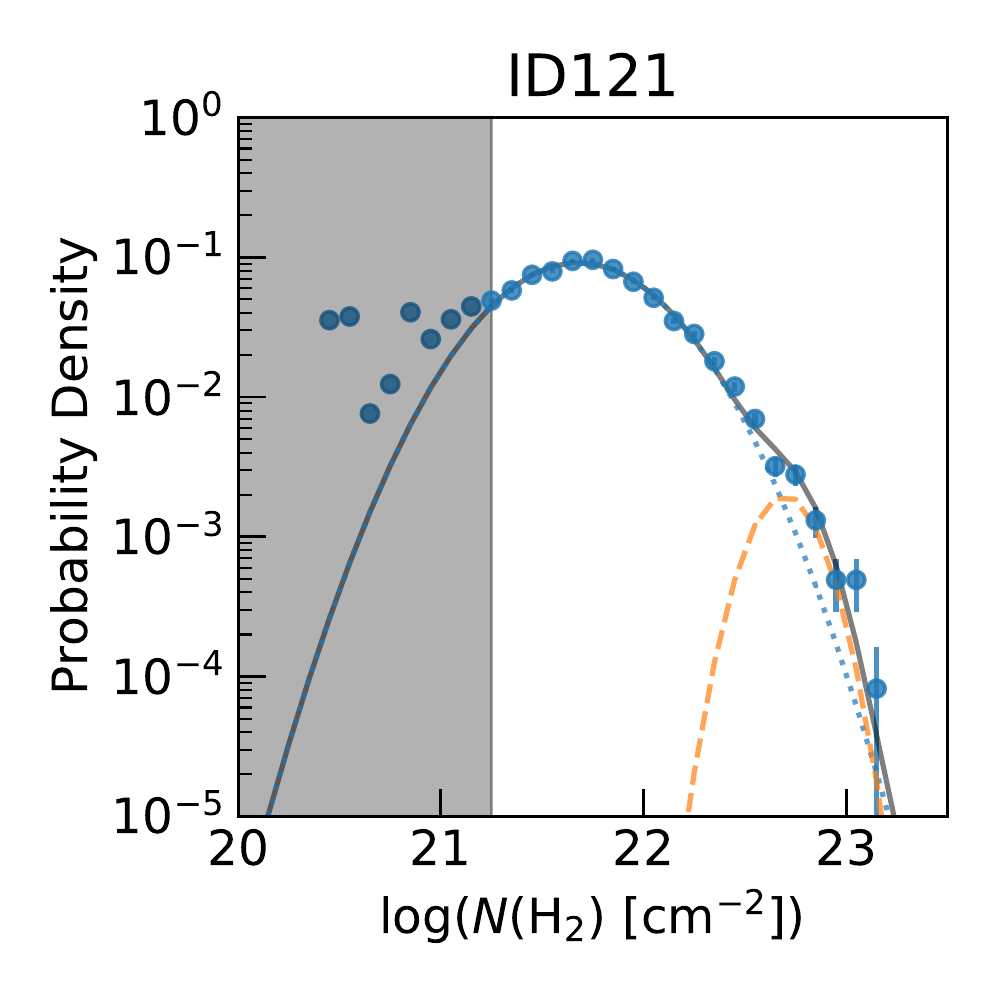}
\includegraphics[width=0.33\textwidth, trim={0 15 0 10}, clip]{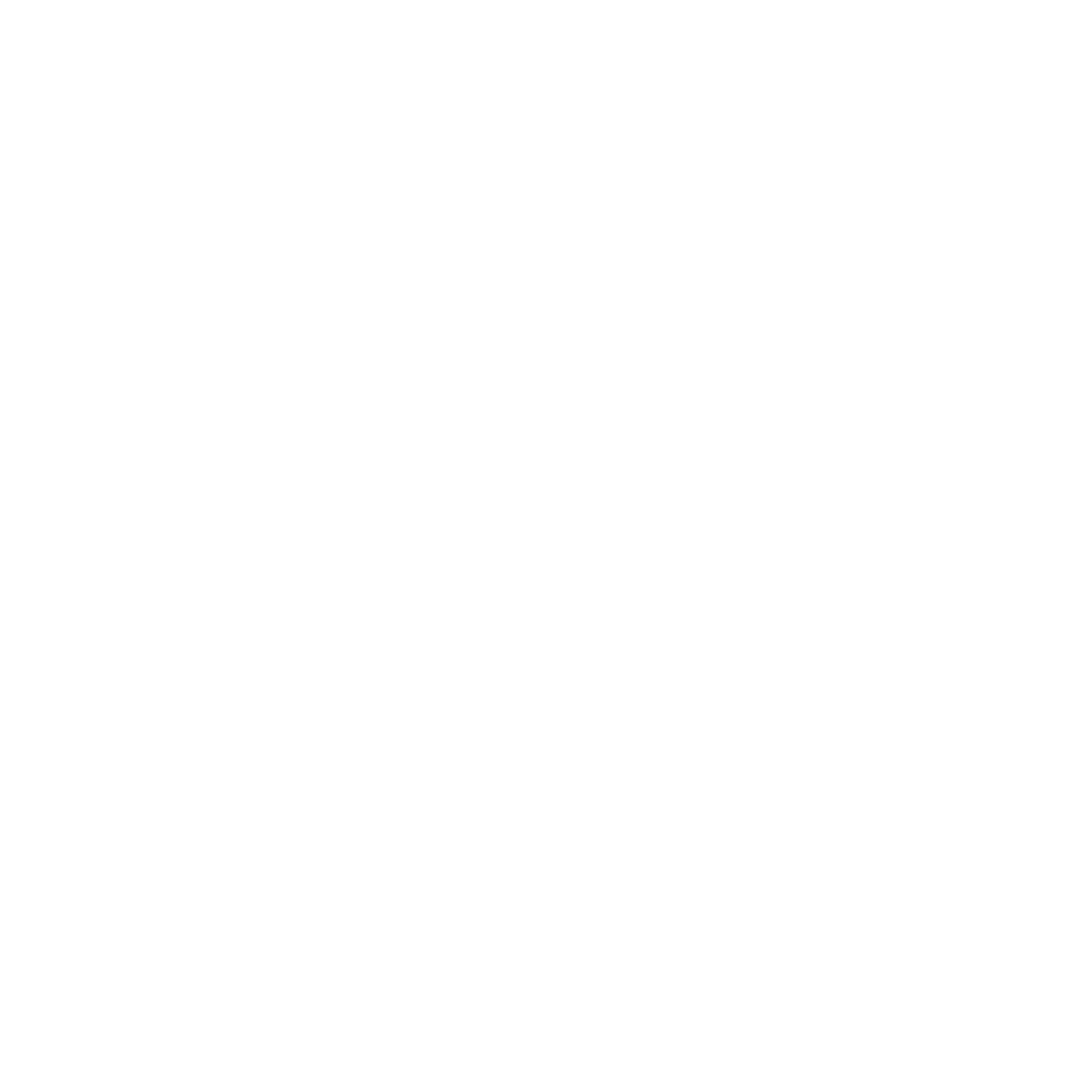}
    \caption{N-PDFs and their best-fit curves for 11 clouds. The blue dotted curves and orange dashed curves indicate the low-density and high-density components, respectively. The black curve is the sum of the two log-normal distributions. The error bars were estimated assuming Poisson noise in each bin. The grey shaded area corresponds to column densities out of the completeness limit.}
    \label{fig:N-PDFs}
\end{figure*}

We found that some of the N-PDFs of our clouds had excess from a single log-normal distribution. Although the densest plots exhibited a steeper decline, the excess component of the N-PDFs was not straight but had a curvature on the log-log plane (Figure \ref{fig:N-PDFs}). In this study, we fitted the N-PDFs with multi-log-normal distributions. The N-PDFs were fitted to the following equation:
\begin{equation}
\label{eq:fit_function}
  p(N) 
  = \sum_{i=1} \frac{p_i}{\sqrt{2 \pi}\sigma_i} \exp \left( 
  \frac{- (\log N - \log \bar{N}_i)^2}{2 \sigma^2_i} \right),
\end{equation}
where $p_i$ and $\bar{N}_i$ are the peak probability density (PD) and the mean column density of each log-normal distribution, respectively.
The first component ($i$ = 1) is the lowest density log-normal distribution, and the following components ($i$ > 1) correspond to higher column density log-normal distributions.
We first performed the fitting using a single log-normal and then increased the log-normal components for molecular clouds with significant residuals.
All parameters of the N-PDFs were fitted using the least-squares method (Levenberg Marquardt algorithm) from the \texttt{astropy.modeling} module for Python \citep{astropy_2013,astropy_2018}.
The best-fit curves and parameters of all N-PDFs of the clouds are shown in Figure \ref{fig:N-PDFs} and Table \ref{tab:N-PDF_params}. 

We found that the N-PDF could be well-fitted using one or two log-normal distributions (Figure \ref{fig:N-PDFs}). Low-density and high-density log-normal distributions were referred to as LN1 and LN2, respectively. In this nonlinear fitting analysis, we set the initial value of the parameter $\bar{N}_1$ to be around the peak value of the low-density component and that of $\bar{N}_2$ to be log($N$(H$_2$)) = 22.5. We confirmed that the fitting result was insensitive to the initial values of $\bar{N}_1$ and $\bar{N}_2$. The column density map of each cloud indicates that the size of the high column density region (log($N$(H$_2$)) > 22.5), which mainly forms LN2, was $\sim 1 - 5$  pc (see Figures \ref{fig:ex_columndensity_1} and \ref{fig:other_columndensity_maps}).

\begin{table*}
    \centering
    \caption{\label{tab:N-PDF_params} Cloud properties and N-PDF fitting parameters of all selected clouds.}
    \begin{threeparttable}
    \begin{tabular}{cccccccccccc}
    \hline
        ID & 
        $\tau$(\cothirteen)\tnote{a} &
        \begin{tabular}[c]{@{}c@{}}$T_\mathrm{ex, median}$\\ {[} K {]}\end{tabular} &
        N(\coeighteen\ clump)\tnote{b} &
        $p_1$  & $\bar{N}_1$  & $\sigma_1$ & $p_2$  & $\bar{N}_2$  & $\sigma_2$ & region & Edge\\
        \hline \hline
        7  & 0.325 & 14.94 & 32 & 0.051 & 21.57 & 0.352 & 0.032 & 22.25 & 0.194 & South & Y \\
        22 & 0.382 & 12.64 & 10 & 0.085 & 21.60 & 0.268 & --    & --    & --    & South & N  \\
        40 & 0.417 & 12.41 & 2  & 0.080 & 21.62 & 0.312 & < 0.001 & 22.60 & 0.148 & South & N   \\
        54 & 0.469 & 11.62 & 28 & 0.085 & 21.65 & 0.316 & --    & --    & --    &  South & Y  \\
        65 & 0.394 & 13.32 & 5  & 0.076 & 21.51 & 0.353 & 0.009 & 22.12 & 0.225 &  South & N  \\
        67 & 0.373 & 10.55 & 2  & 0.080 & 21.45 & 0.322 & --    & --    & --    &  South & Y  \\
        95 & 0.288 & 13.62 & 7  & 0.084 & 21.54 & 0.490 & --    & --    & --    &  North & N  \\
        96 & 0.384 & 10.74 & 9  & 0.073 & 21.66 & 0.211 & --    & --    & --    &  North & Y  \\
        107& 0.362 & 12.06 & 13 & 0.074 & 21.46 & 0.302 & 0.013 & 22.17 & 0.261 &  North & N  \\
        119& 0.304 & 16.16 & 19 & --    & --    & --    & --    & --    & --    &  North & Y   \\
        121& 0.314 & 14.44 & 14 & 0.082 & 21.68 & 0.359 & 0.001 & 22.70 & 0.148 &  North & Y  \\
        \hline
    \end{tabular}
    \begin{tablenotes}
        \item[a] The median optical depth from \cothirteen\ data.
        \item[b] The \coeighteen\ clumps were identified by \cite{Takekoshi2019}.
    \end{tablenotes}
    \end{threeparttable}
\end{table*}

\section{Discussion}
\label{sec:discussion}

\subsection{Absence of power-law distribution}
\label{sec:absence_pl}
In this section, we discussed the observed N-PDF shapes. Why do our clouds not present a power-law tail? We focused on the size of structures that exhibit power-law tails.

\cite{Ma2021} obtained the N-PDFs for 40 molecular clouds identified in the Milky Way Imaging Scroll Painting (MWISP) project \citep{Su2019}. They reported that power-law tails are formed from structures approximately 1 pc in size. In contrast, 
\cite{Chen2018} reported that the power-law tail of the N-PDF was the sum of the N-PDFs of the sub-structures inside the molecular cloud. The spatial size of these sub-structures was approximately 0.5 pc. This structure size which formed a power-law tail is also supported by numerical studies \citep[e.g.][]{Chen2018,Khullar2021}. We can consider that the size of the structure forming the power-law component is less than 1 pc from previous N-PDF studies. Here, it should be noted that the N-PDF analysis requires a pixel resolution of one order of magnitude better than the size of the structure to be discussed because each N-PDF requires approximately $10^2$ pixels. For dense and compact structures that form power-law tails to affect statistics, such as PDF, we consider that a spatial resolution of 0.01 pc is required. Therefore, the spatial resolution of the Cygnus X Survey data ($\sim$ 0.3 pc) is too poor to discuss the spatial structures and their background mechanism in the high-density regime of the N-PDF. 

In this analysis, it is impressive that the N-PDFs obtained from the molecular gas tracer could be well-fitted with one or two log-normal distributions. This means that the ``power-law'' component reported in previous observational studies could be the second component of our study. This result suggests an alternative density structure of molecular clouds based on the conventional picture (see Section \ref{sec:hierarchical_density_structure}).

\subsection{Relation between N-PDF and star-forming activity}
\label{sec:N-PDF_SF_activity}

\begin{figure*}
\includegraphics[width=0.45\textwidth, trim={10 35 10 50}, clip]{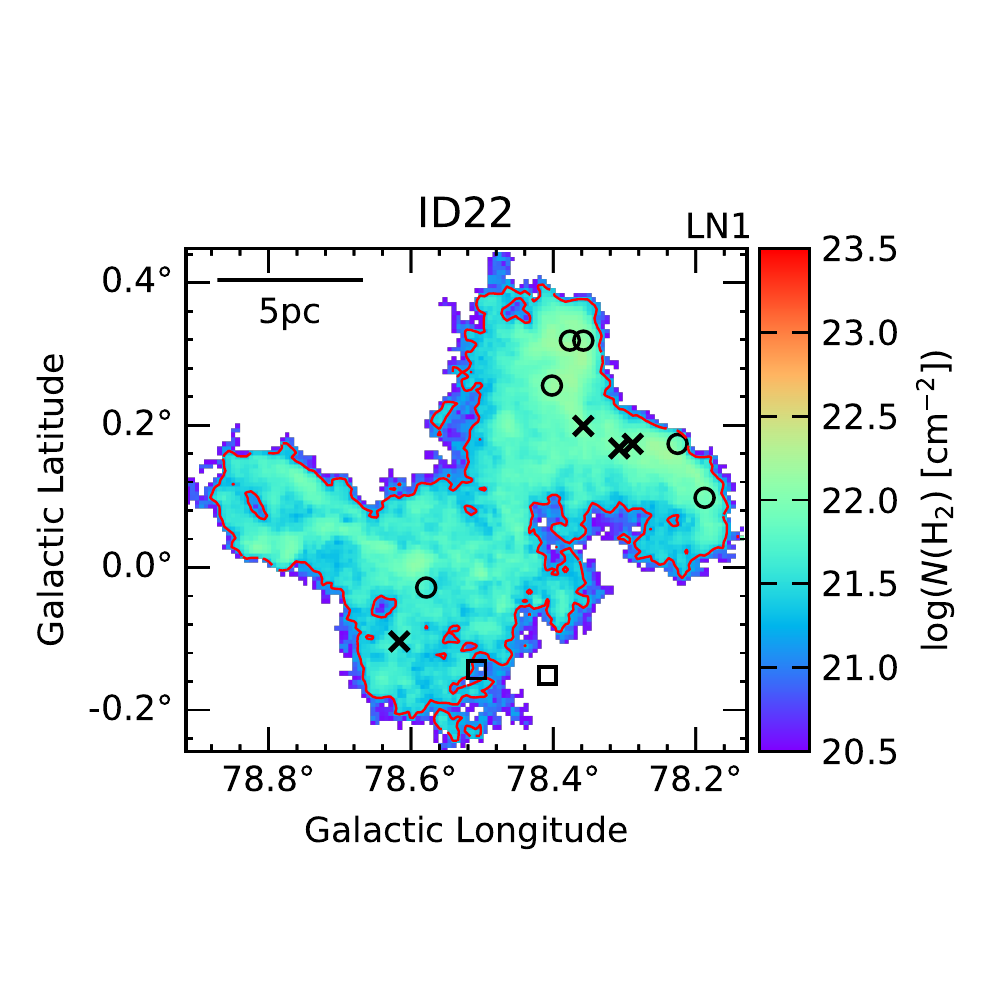}
\includegraphics[width=0.45\textwidth, trim={0 33 10 40}, clip]{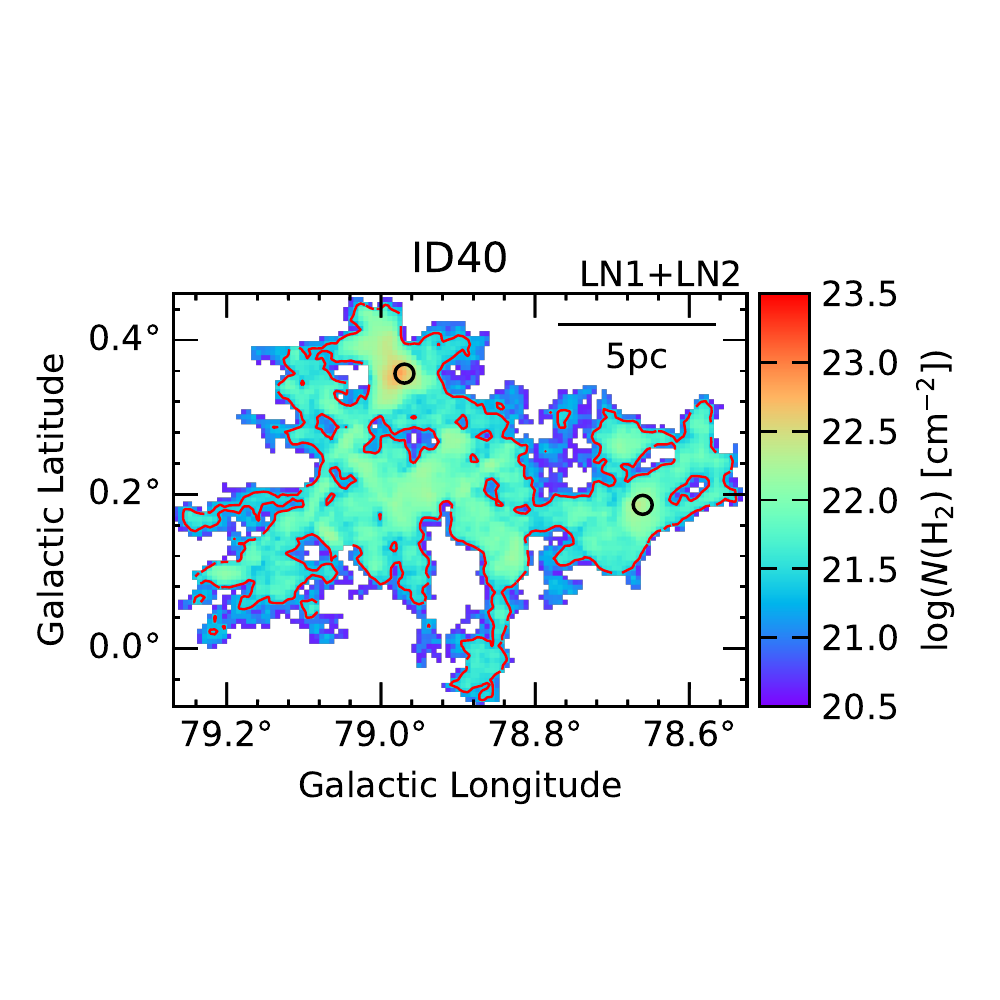}
\includegraphics[width=0.45\textwidth, trim={10 35 10 40}, clip]{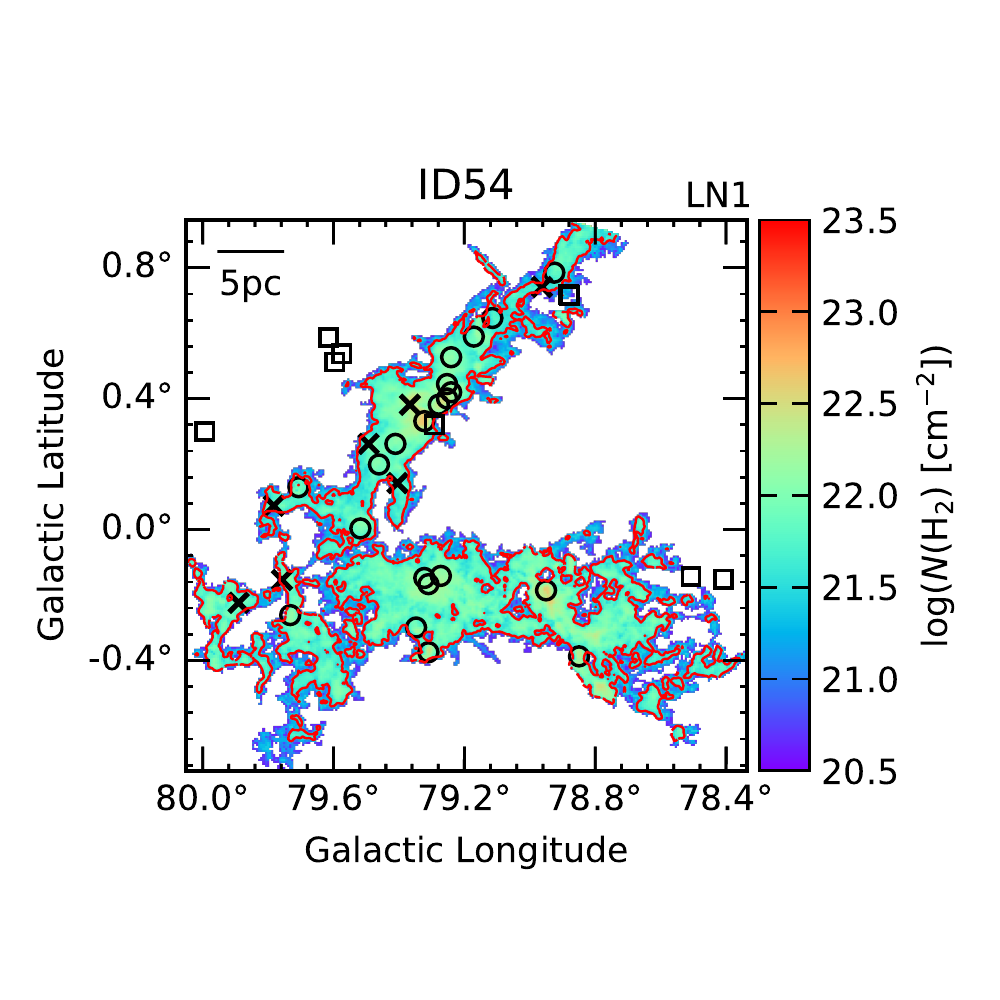}
\includegraphics[width=0.45\textwidth, trim={0 30 10 40}, clip]{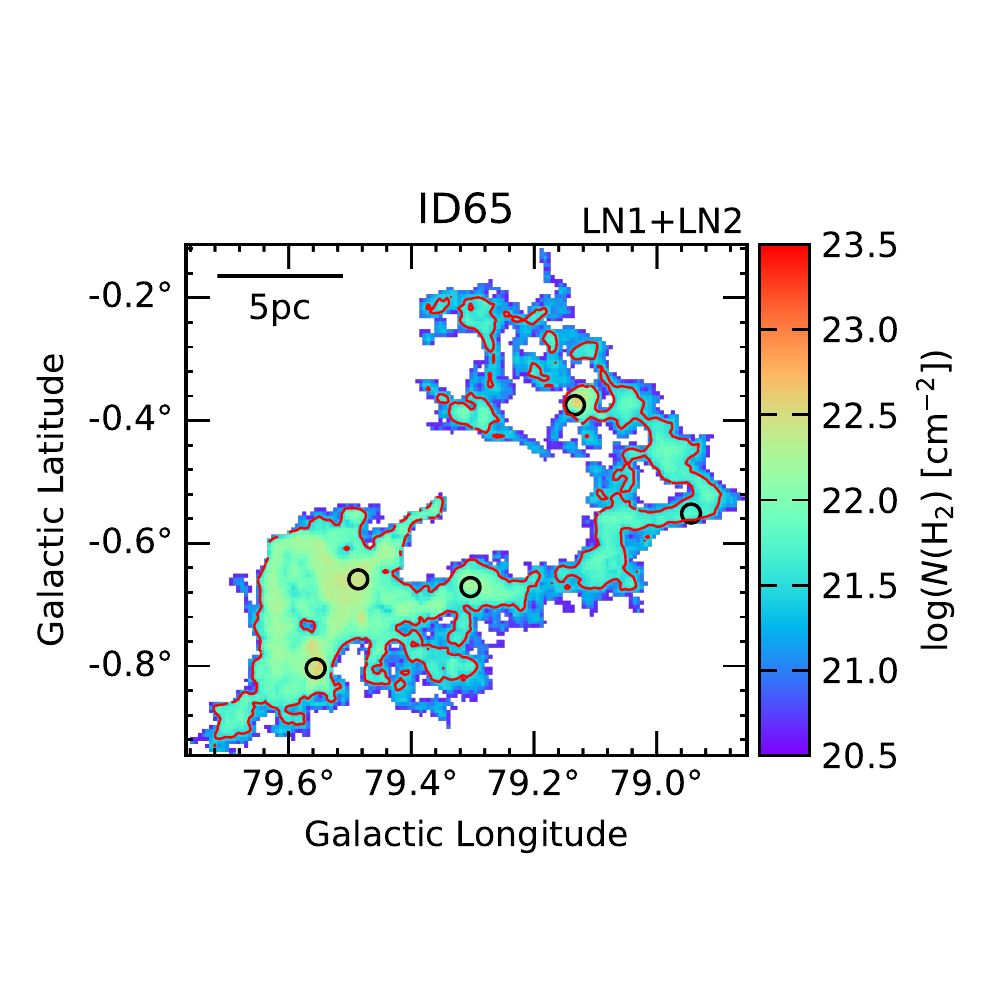}
\includegraphics[width=0.45\textwidth, trim={10 20 10 10}, clip]{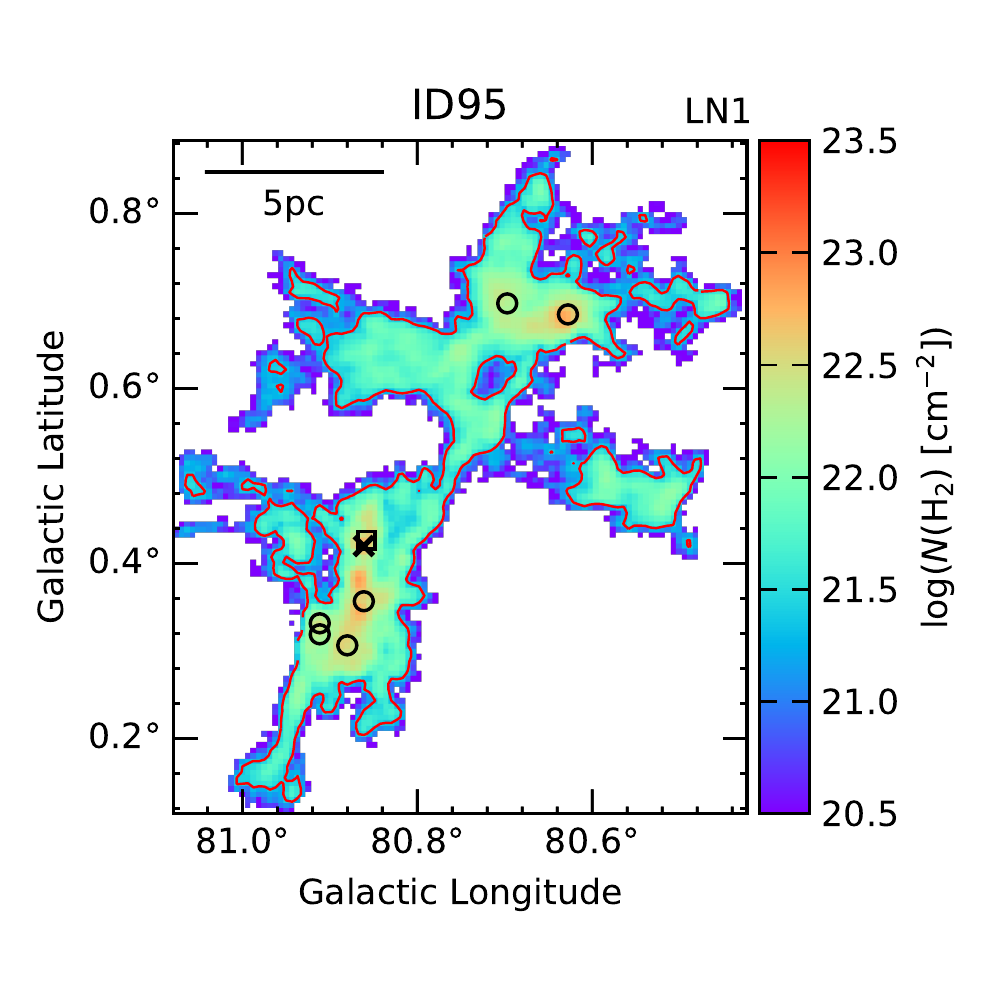}
\includegraphics[width=0.45\textwidth, trim={0 18 10 10}, clip]{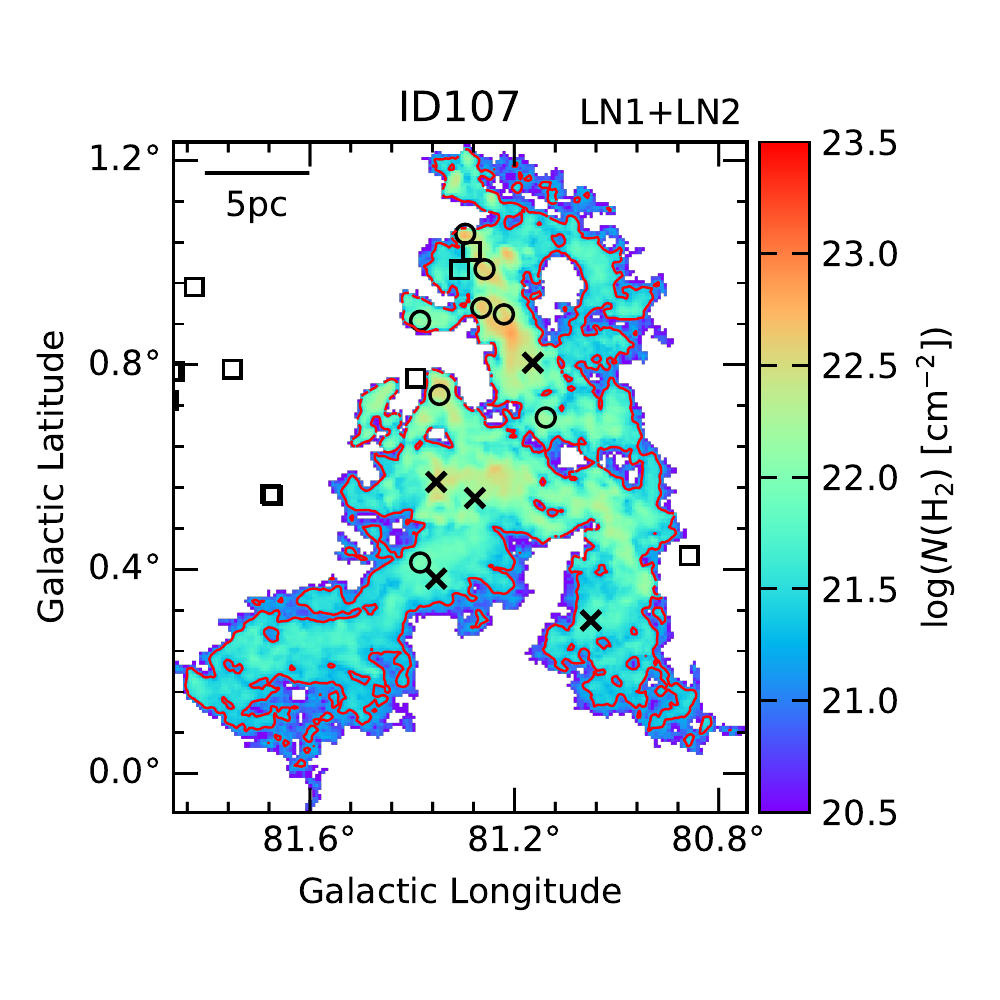}
    \caption{H$_2$ column density maps of some clouds in our sample. The red contour indicates the completeness limit. The circles and crosses indicate the positions of the prestellar and starless cores identified by \citet{Takekoshi2019} from \coeighteen\ intensity maps, respectively. The square marks indicate the centre positions of the radio continuum sources obtained by \citet{Urquhart2009}. The other column density maps can be found in Figure \ref{fig:other_columndensity_maps}.}
    \label{fig:ex_columndensity_1}
\end{figure*}

Previous studies have suggested that the shape of the N-PDF is related to its star-forming activity in the interior of the molecular cloud \citep[e.g.][]{Kainulainen2009,Pineda2010,Ballesteros-Paredes2011,Schneider2013,Tremblin2014,Lombardi2015,Ma2021,Lewis2022}. In this section, we discussed the shape of the N-PDFs and star-forming activity using molecular clouds with mainly closed boundaries. We investigated the relationship between the shape of the N-PDFs and star-forming activity by comparing the column density and spatial distributions of high-density clumps and radio continuum sources in each cloud. The following three cases were considered:

Figures \ref{fig:ex_columndensity_1} and \ref{fig:other_columndensity_maps} show the centre positions of the \coeighteen\ clumps identified by \citet{Takekoshi2019}, superimposed on each column density map. These clumps were classified as prestellar and starless cores, which are indicated by circles and crosses in the figures, respectively. Column 4 in Table \ref{tab:N-PDF_params} shows the number of associated \coeighteen\ clumps for each cloud. We found no simple correlation between the number of \coeighteen\ clumps associated with molecular clouds and the shape of the N-PDF. The ID40 and ID65 clouds had an excess component in the high-density range. However, these clouds had a few \coeighteen\ clumps. In contrast, the ID22 and ID54 clouds had more \coeighteen\ clumps than the ID40 and ID65 clouds, although its N-PDF had a single log-normal distribution.

Next, we investigated the relationship between the N-PDF and star-formation activity in each molecular cloud. In addition to the positions of the \coeighteen\ clumps, the positions of the continuum sources identified by \citet{Urquhart2009} are shown on each column density map as indicators of high-mass star formation (square marks in Figures \ref{fig:ex_columndensity_1} and \ref{fig:other_columndensity_maps}). These continuum sources have been classified as compact and ultra-compact \HII regions. The ID40, ID65, and ID107 clouds shown in Figure \ref{fig:ex_columndensity_1} indicate LN1 plus LN2 shape. However, there were no continuum sources associated with the clouds at ID40 and ID65. In addition, the ID107 cloud shows a poor positional correlation between the high-density region and the continuum sources. It suggests that both clouds have only a weak correlation between high-density regions and star-forming activity. The reverse cases are found. The ID95 cloud showed only a single log-normal N-PDF, even though the continuum sources were located in high-density regions. The ID54 cloud, in which many prestellar cores are located, also exhibited a single log-normal. Therefore, we suggest that there is little relationship between the star formation activity and N-PDF.

We then investigated the relationship between the star formation stages of the molecular cloud complexes and the N-PDFs of the molecular clouds contained within them. We found no correlation between the star formation stage of the molecular cloud complexes and the density structures of the molecular clouds. As mentioned in Section \ref{sec:cyg-x}, the Cygnus X complex has been reported to exhibit different star-forming stages in the northern and southern regions. Table \ref{tab:N-PDF_params} presents a molecular cloud complex and the N-PDF parameters to which the identified cloud belongs. In this study, the number of N-PDFs with excess components was almost the same in both regions. In addition, the log-normal N-PDF parameters (e.g. mean column density and width of log-normal distribution) of the clouds obtained in Section \ref{sec:fit_PDF} exhibited no prominent differences in both northern and southern complexes (see Section \ref{sec:hierarchical_density_structure} for details).

In conclusion in Section \ref{sec:N-PDF_SF_activity}, we found that the shape of the N-PDF is not necessarily related to star formation activity. Although more investigations are needed to obtain more robust results, we consider the N-PDF reflects the relationship between the density structure and the turbulent structure of the molecular cloud.

\subsection{Multiple density structure in molecular clouds}
\label{sec:hierarchical_density_structure}
\begin{figure}
    \includegraphics[width=\linewidth, trim={0 0 0 0}, clip]{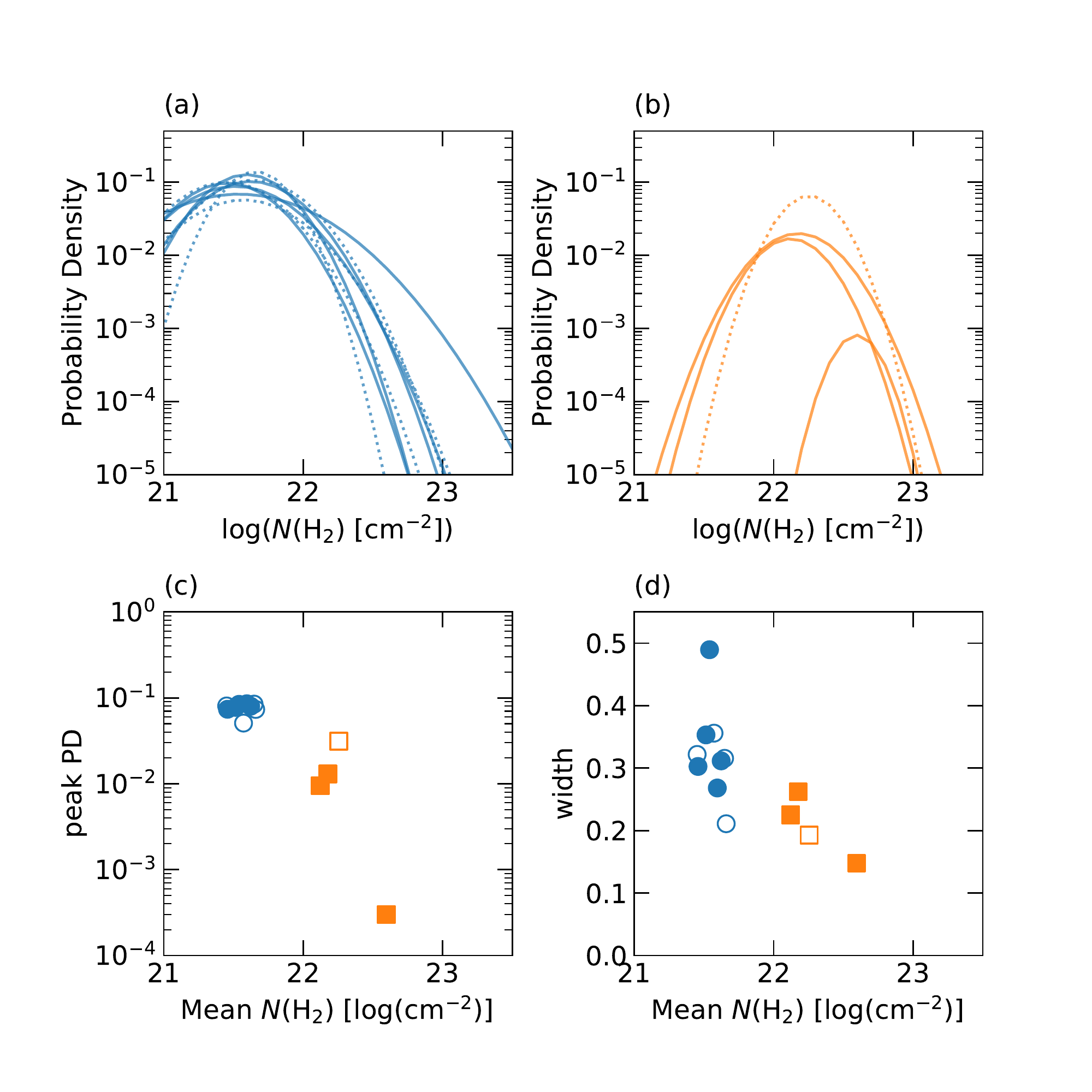}
    \caption{Fitting results of N-PDFs of our clouds. 
    Panels (a) and (b) indicate the best-fit curves for the LN1 and LN2 shown in Figure \ref{fig:N-PDFs}, respectively. 
    The solid lines and dotted lines indicate the completeness limit contours were closed and the other clouds, respectively. 
    Panels (c) and (d): correlation diagrams between the mean column density and the peak PD of the cloud, and (d) between the mean column density and width of LN components, respectively. The peak PD value of each LN component is $p_i$ in equation (6).
    The blue circles and orange squares indicate LN1 and LN2, respectively. The filled symbols indicate the clouds of which the completeness limit contours were closed, and the open symbols indicate the other clouds.}
    \label{fig:cor_fit_gauss}
\end{figure}

We performed a statistical investigation of the N-PDF using the fitting parameters of the N-PDF shown in Section \ref{sec:fit_PDF}.
In this study, we focused on the correlation between the peak PD, $p_i$ and the width, $\sigma_i$ of each log-normal distribution with the mean column density, $\bar{N}_i$.

Panels (a) and (b) of Figure \ref{fig:cor_fit_gauss} show the best-fit curves for LN1 and LN2, respectively, obtained from our clouds.
In these panels, we found that ten clouds had peak column densities of $\sim$ 10$^{21.5}$ \scm\ for LN1; LN2 showed different curves.
This suggests that the physical properties of LN1 are universal and independent of the molecular cloud characteristics.
Panels (c) and (d) of Figure \ref{fig:cor_fit_gauss} show the correlation plots of the N-PDF parameters. 
Circles and squares indicate LN1 and LN2, respectively.
To evaluate the effect of the completeness limit, we used closed and open symbols for closed and unclosed contours at the completeness limit, respectively (see Section \ref{sec:columndesity}).
Figure \ref{fig:cor_fit_gauss}-(c) shows the correlation between the mean column density and the peak PD of the log-normal components for each cloud.
We found that the higher mean column density shows a smaller peak PD.  
This is consistent with the hierarchical spatial and density structures in a molecular cloud \citep[e.g.][]{Larson1994,Stutzki1998,McKee2007}. 
Figure \ref{fig:cor_fit_gauss}-(d) shows the correlations between the mean column density and the widths of LN1 and LN2.
All plots of LN1 show a large dispersion by a factor of 2.3 (from 0.211 to 0.490) along the width axis. 
The plots for the closed contour clouds show a similar dispersion by a factor of 1.83 within a 0.16 dex range of the peak column density.
In contrast, the plots of LN2 show systematically smaller widths than LN1, which were narrower than 0.27.
Through both components, we found a systematic correlation: a component with the larger peak density had a narrower width. 

The two correlations shown in Figure \ref{fig:cor_fit_gauss} suggest that a higher density component has the smaller volume fraction and the smaller turbulence velocity within each structure.
Previous observational studies have reported that interstellar matter has a hierarchical structure consisting of small dense regions within a less dense and larger envelope.
It is not a single hierarchical level, but many layers over a wide range of scales from kpc to sub-pc \citep[e.g.][]{Falgarone1991,Blitz1993,Larson1994,Armstrong1995,Larson1995,Stutzki1998,Elmegreen2000,Williams2000,Elmegreen2004,Chepurnov2010}.
In addition to this spatial hierarchical structure, we found a turbulence hierarchical structure at the molecular cloud scale (Figures \ref{fig:N-PDFs} and \ref{fig:cor_fit_gauss}).

What is the origin of these log-normal components? 
The decrease in turbulent velocity shown in Figure \ref{fig:cor_fit_gauss}-(d) can be considered as a result of the supersonic turbulence cascade in the ISM \citep[e.g.][]{Larson1981,Kritsuk2013}.
However, numerical studies resolving from the supersonic to subsonic scales have reported that the gas density PDF exhibits a single log-normal \citep{Federrath2021}.
To form the multiple log-normal density structure, it will be necessary to consider origins other than turbulent cascades.

Numerical simulations have been performed to investigate the thermal evolution of a multiphase ISM with supersonic turbulence.
\cite{Kobayashi2022} found that the shocked inhomogeneous ISM exhibits density distribution functions composed of several log-normal distributions with different peak densities and widths corresponding to different temperature ranges \citep[see also Figure 4 in][]{Kobayashi2022}.
The spatial structure induced by the thermal instability has also been reported.
It has dense cold clumps with slow turbulent velocity embedded in a diffuse warm interclump gas dominated by stronger turbulence.
It should be noted, however, that the density and temperature ranges observed in this study were different from the numerical simulations, such that \cite{Kobayashi2022}.
They focused on molecular clouds in the early evolutional stages, when \HI gas was still dominant.

As noted in Section \ref{sec:absence_pl}, it is unclear why the density structures of molecular clouds can be fitted only by two log-normal distributions. To address this issue, observational studies of molecular clouds in different environments and numerical simulations of the multiphase ISM at later evolutional stages are required.

\section{Conclusions}
\label{sec:conclusion}
We investigated the N-PDF based on molecular emission lines using the Nobeyama 45-m Cygnus X CO survey data. Using the DENDROGRAM and SCIMES algorithms, molecular clouds were identified, and the N-PDF properties were investigated for extended clouds (> 0.4 deg$^2$). The main results of this study are as follows:

\begin{enumerate}
    \item We found that the N-PDFs obtained from ten out of the eleven selected clouds could be well-fitted with one or two log-normal distributions. 
    \item We found that the shape of the N-PDF in cloud scale had less correlation with the star-forming activity in the interior of the molecular cloud. 
    Although, it should be kept in mind that star formation activity in the interior of molecular clouds has a significant impact on the CO abundance, especially for low-$J$ transitions of CO isotopes.
    \item From the fitting of the N-PDFs, we found that LN1 had almost the same mean column density ($\sim$ 10$^{21.5}$ \scm) regardless of the features of the individual molecular cloud.
    \item We found the correlations between the parameters which characterise the N-PDFs. The width of the log-normal distribution decreased as the mean column density of the structure increased. This tendency implies that the turbulent velocity inside the structure is smaller when the mean column density increases.
\end{enumerate}

Our fitting of the multi-log-normal distributions for the N-PDF revealed an unconventional molecular cloud density structure. However, the reason why the density structures of the molecular clouds are two log-normal but not three or more log-normals remains a question. Further investigations of the physical processes are required to explain the results.

Although the statistical results are not robust, the log-normal distribution in the low-density part (LN1) exhibited common physical properties, regardless of the molecular cloud characteristics. The LN2 indicated different properties depending on the molecular cloud. These results are interesting for studying the scale from molecular clouds to star-forming cores. We are extending our approach to large-scale maps at different spatial resolutions (e.g. FUGIN) to determine the statistical physical properties of molecular clouds and the results will soon appear.

\section*{Acknowledgements}
The 45-m radio telescope is operated by the Nobeyama Radio Observatory, a branch of the National Astronomical Observatory of Japan. This research used \textsc{aplpy}, an open-source plotting package for Python \citep{aplpy2012}, \textsc{astrodendro},\footnote{\url{http://www.dendrograms.org/}} a Python package to compute dendrograms of Astronomical data \citep{Rosolowsky2008}, \textsc{astropy},\footnote{\url{http://www.astropy.org}} a community-developed core Python package for Astronomy \citep{astropy_2013, astropy_2018}, \textsc{matplotlib}, a Python package for visualisation \citep{Hunter_2007}, \textsc{numpy}, a Python package for scientific computing \citep{harris_2020}, \textsc{scimes}, a Python package for finding relevant structures in dendrograms of molecular gas emission using the spectral clustering approach \citep{Colombo2015}, and Overleaf, a collaborative tool. We would like to thank Editage (\url{www.editage.jp}) for English language editing.


\section*{Data Availability}
The data underlying this article are available in the article.
 



\bibliographystyle{mnras}
\bibliography{murase_2022a} 




\appendix
\section{Column density map of our clouds}
\begin{figure*}
\includegraphics[width=0.45\textwidth, trim={10 20 0 20}, clip]{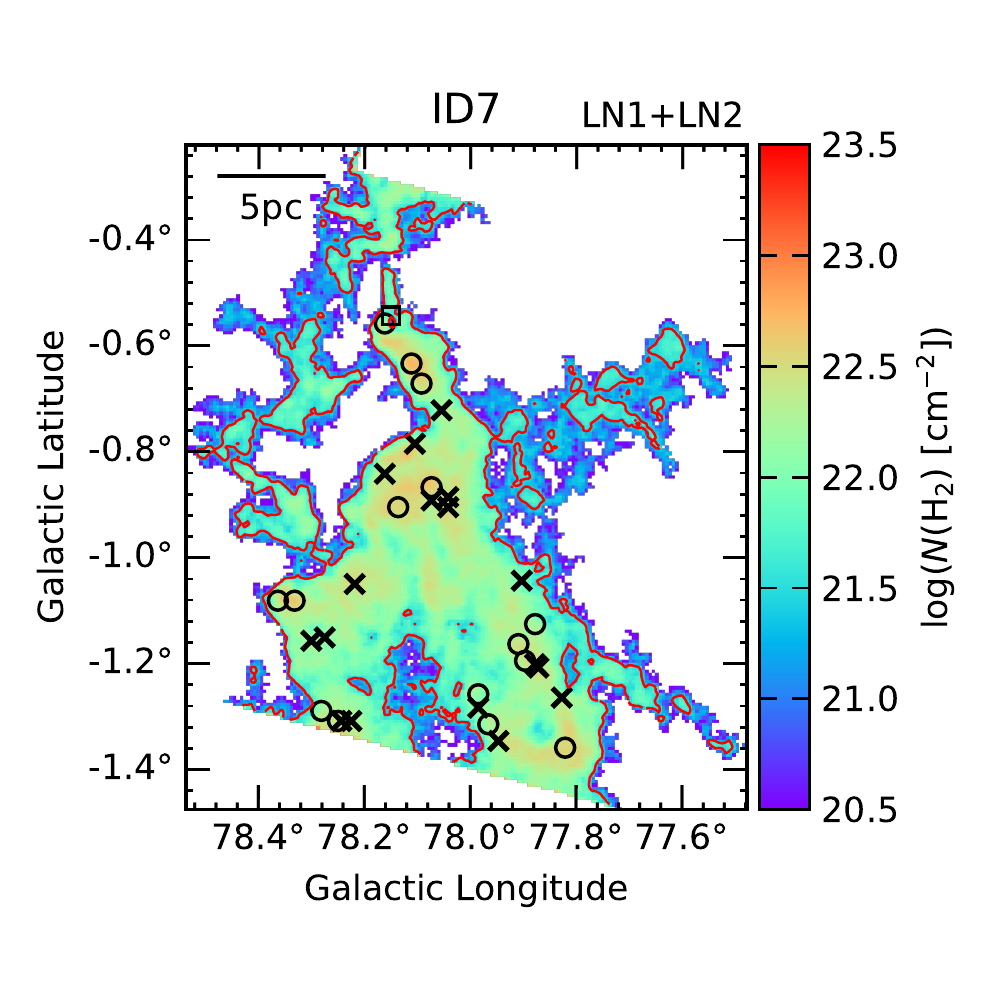}
\includegraphics[width=0.45\textwidth, trim={10 18 0 40}, clip]{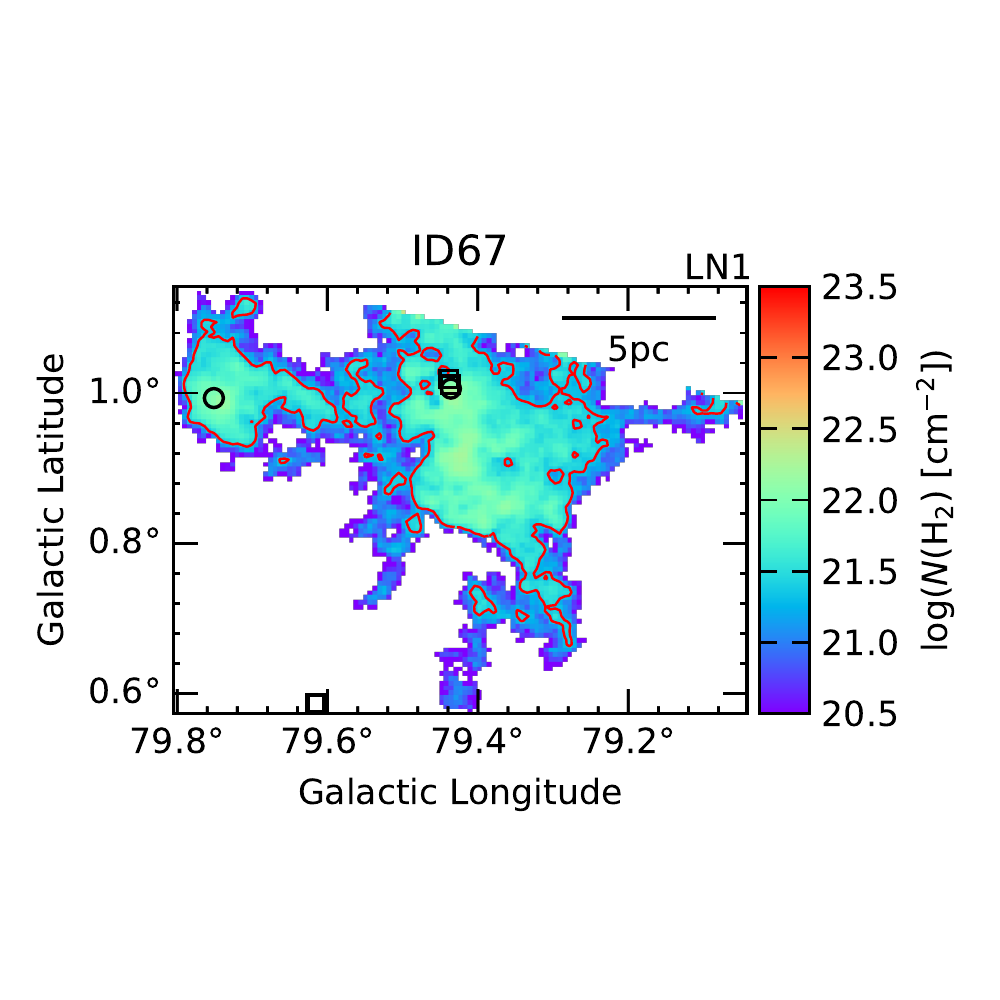}
\includegraphics[width=0.45\textwidth, trim={10 0 0 50}, clip]{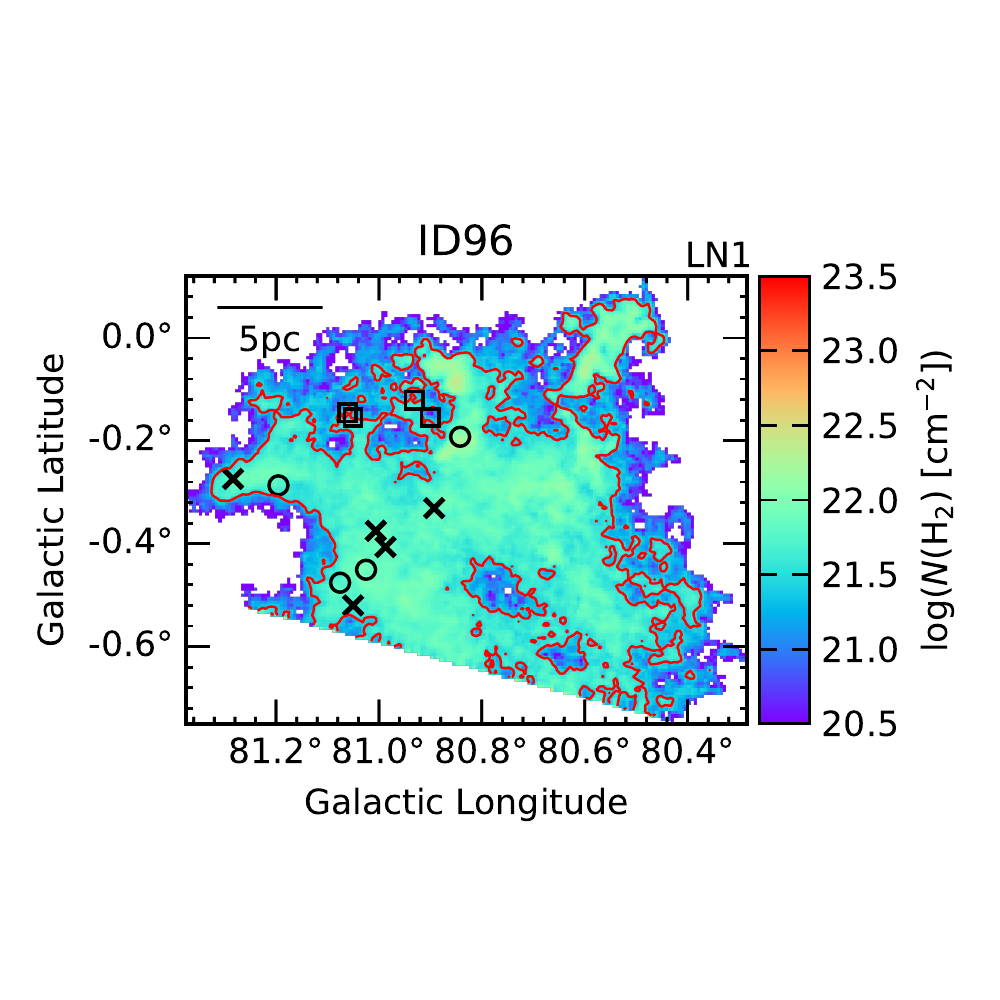}
\includegraphics[width=0.45\textwidth, trim={0 10 10 0}, clip]{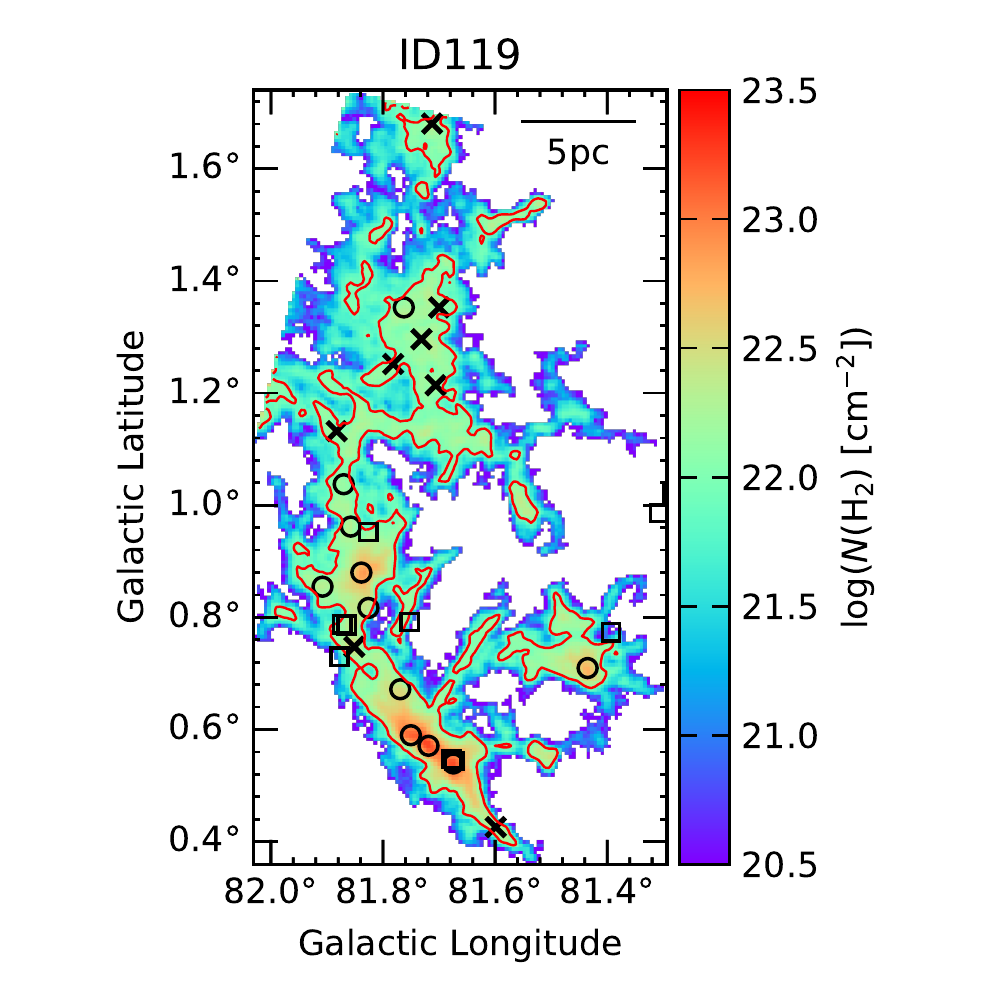}
\includegraphics[width=0.45\textwidth, trim={10 25 0 25}, clip]{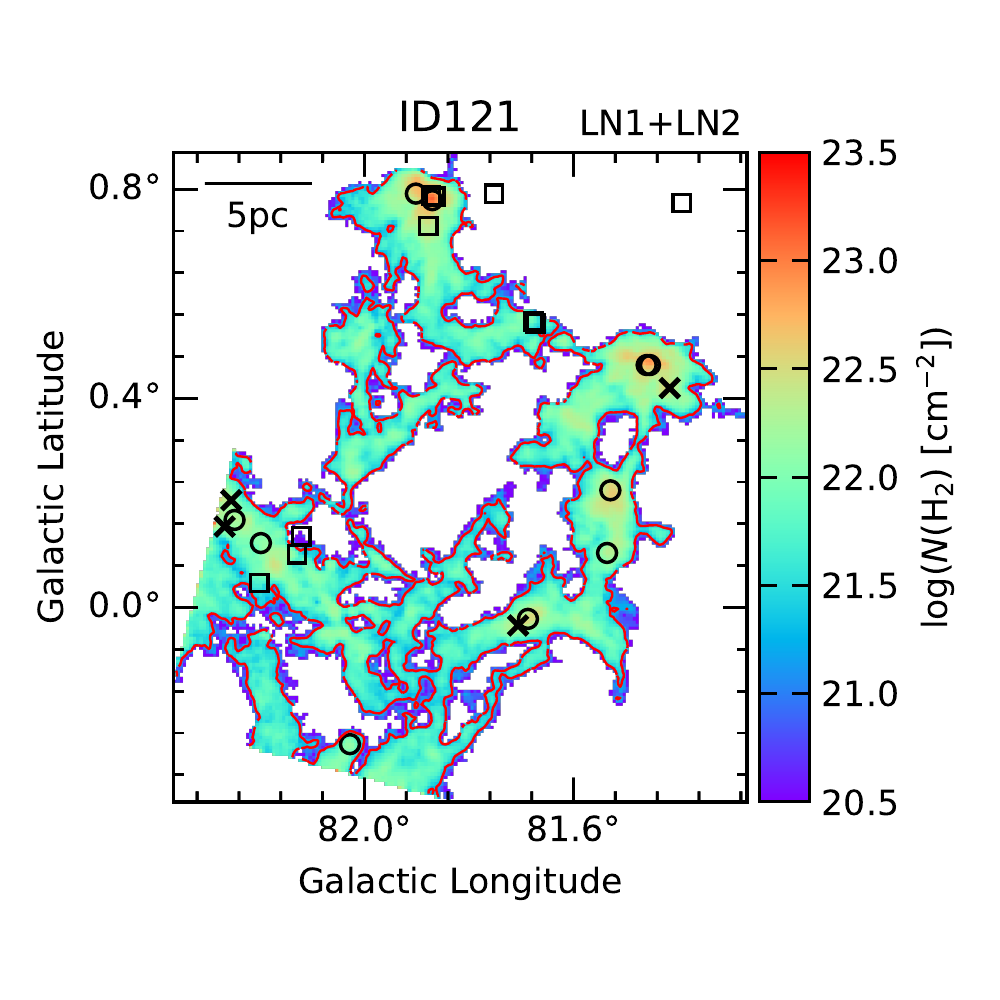}
\includegraphics[width=0.45\textwidth, trim={0 25 0 30}, clip]{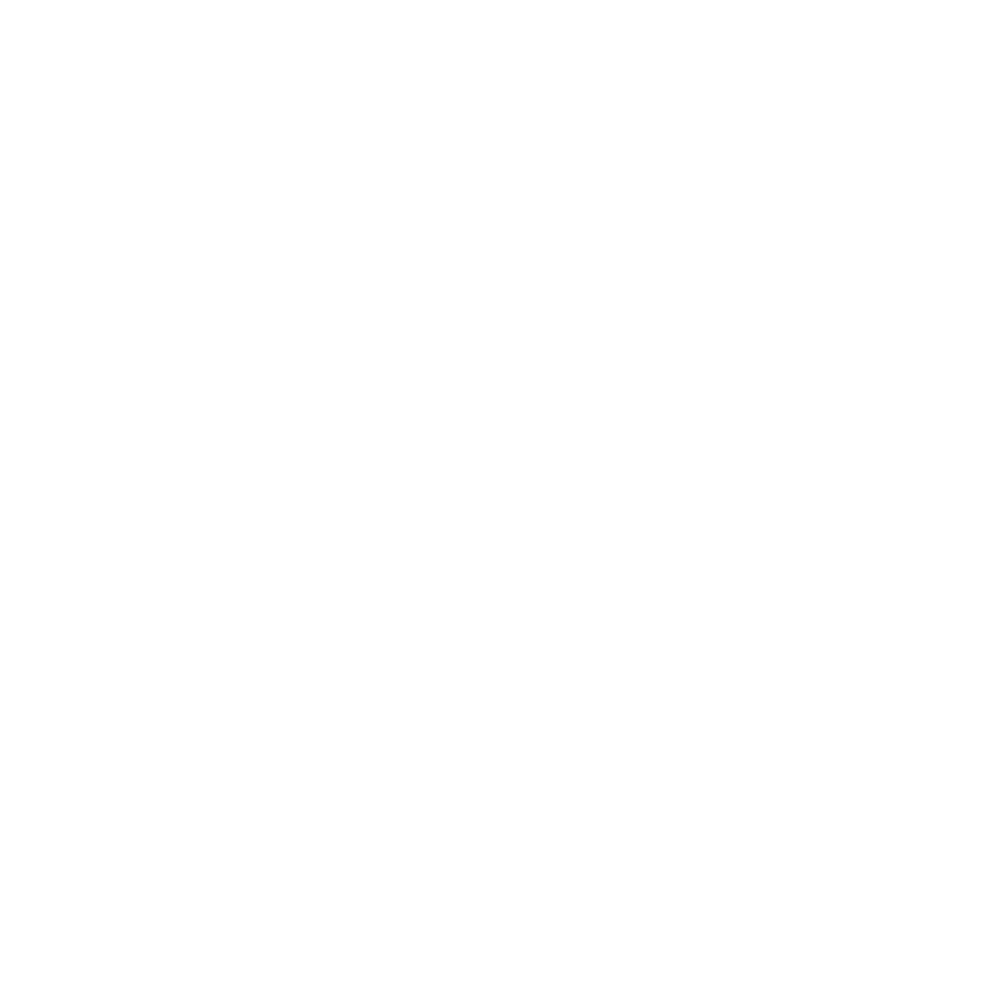}
    \caption{Column density maps of other selected clouds with Figure \ref{fig:ex_columndensity_1}. The red contours and marks are the same as in Figure \ref{fig:ex_columndensity_1}.}
    \label{fig:other_columndensity_maps}
\end{figure*}

\clearpage
\section{Online material}
\begin{table}
    \centering
    \caption{All cloud parameters identified using SCIMES.}
    \label{table:cloud_property_online}
    \begin{tabular}{cccccc}
    \hline
        ID   &
        \begin{tabular}[c]{@{}c@{}}$l$\\ {[} deg. {]}\end{tabular} & \begin{tabular}[c]{@{}c@{}}$b$\\ {[} deg. {]}\end{tabular} & \begin{tabular}[c]{@{}c@{}}$v_\mathrm{LSR}$\\ {[} \kms {]}\end{tabular} & \begin{tabular}[c]{@{}c@{}}$\sigma_v$\\ {[} \kms {]}\end{tabular} &
        \begin{tabular}[c]{@{}c@{}}$S_\mathrm{exact}$\\ {[} pix.$^2$ {]}\end{tabular}  \\
    \hline \hline
        1   & 77.874 & $-$0.477 & $-$1.85 & 0.74  & 101   \\
        2   & 77.882 & $-$1.270 & $-$7.51 & 1.63  & 446   \\
        3   & 77.964 & $-$0.437 & $-$5.32 & 0.89  & 317   \\
        4   & 77.981 & $-$0.397 & 11.53 & 0.68  & 142   \\
        5   & 78.003 & $-$1.064 & 13.75 & 1.03  & 402   \\
        6   & 78.058 & $-$0.459 & $-$3.33 & 0.55  & 226   \\
        7   & 78.063 & $-$1.009 & $-$0.15 & 2.34  & 14202 \\
        8   & 78.065 & $-$0.381 & 13.25 & 0.34  & 68    \\
        9   & 78.122 & $-$0.860 & 14.51 & 0.63  & 296   \\
        10  & 78.157 & $-$1.280 & 14.00 & 1.00  & 485   \\
        11  & 78.166 & 0.002  & 0.73  & 1.06  & 213   \\
        12  & 78.203 & $-$0.148 & $-$7.99 & 1.56  & 102   \\
        13  & 78.218 & $-$0.314 & 13.69 & 0.47  & 357   \\
        14  & 78.226 & 0.536  & $-$0.59 & 2.78  & 3656  \\
        15  & 78.235 & 0.234  & 14.13 & 0.62  & 544   \\
        16  & 78.323 & 0.370  & $-$3.28 & 0.47  & 234   \\
        17  & 78.368 & $-$0.476 & 11.26 & 2.03  & 1870  \\
        18  & 78.403 & $-$1.047 & 12.86 & 1.42  & 478   \\
        19  & 78.408 & $-$0.241 & 13.47 & 0.38  & 139   \\
        20  & 78.438 & 0.395  & $-$0.94 & 1.71  & 379   \\
        21  & 78.475 & 0.066  & $-$6.98 & 1.10  & 310   \\
        22  & 78.479 & 0.105  & 8.63  & 1.30  & 5915  \\
        23  & 78.494 & $-$1.256 & 13.01 & 0.75  & 148   \\
        24  & 78.517 & $-$0.328 & 12.63 & 0.47  & 273   \\
        25  & 78.535 & $-$0.852 & 5.63  & 0.97  & 1499  \\
        26  & 78.553 & 0.179  & 4.11  & 0.55  & 232   \\
        27  & 78.568 & $-$0.149 & $-$6.77 & 1.02  & 82    \\
        28  & 78.572 & $-$1.210 & $-$0.47 & 1.96  & 238   \\
        29  & 78.606 & 0.175  & 1.77  & 0.41  & 200   \\
        30  & 78.632 & 0.263  & 8.70  & 0.55  & 283   \\
        31  & 78.639 & 0.425  & 3.02  & 0.54  & 131   \\
        32  & 78.657 & 0.772  & $-$5.54 & 1.24  & 2475  \\
        33  & 78.667 & $-$1.026 & 5.48  & 0.43  & 264   \\
        34  & 78.703 & $-$0.055 & 2.98  & 0.42  & 459   \\
        35  & 78.762 & $-$1.120 & 4.92  & 0.39  & 217   \\
        36  & 78.777 & $-$1.109 & $-$0.08 & 1.61  & 481   \\
        37  & 78.837 & $-$0.902 & 11.06 & 1.22  & 342   \\
        38  & 78.875 & $-$0.735 & $-$3.03 & 0.77  & 379   \\
        39  & 78.885 & $-$0.615 & 4.59  & 0.34  & 263   \\
        40  & 78.916 & 0.223  & 4.06  & 4.58  & 4492  \\
        41  & 78.929 & $-$0.614 & 8.05  & 1.08  & 335   \\
        42  & 78.993 & 0.866  & 9.20  & 0.42  & 177   \\
        43  & 78.996 & 0.051  & $-$4.18 & 1.31  & 232   \\
        44  & 79.005 & 0.614  & 6.37  & 1.03  & 532   \\
        45  & 79.057 & 0.475  & $-$2.61 & 0.84  & 290   \\
        46  & 79.077 & $-$0.284 & 12.53 & 0.50  & 245   \\
        47  & 79.080 & $-$0.969 & 12.71 & 0.97  & 337   \\
        48  & 79.107 & $-$0.070 & 8.29  & 0.55  & 89    \\
        49  & 79.181 & 0.956  & 7.21  & 0.88  & 1009  \\
        50  & 79.186 & 0.428  & 11.11 & 0.70  & 97    \\
        51  & 79.188 & $-$1.013 & 9.03  & 3.00  & 1053  \\
        52  & 79.195 & 0.534  & 10.23 & 0.64  & 422   \\
        53  & 79.199 & 0.400  & 8.29  & 0.68  & 370   \\
        54  & 79.235 & $-$0.036 & $-$0.16 & 2.92  & 21290 \\
        55  & 79.237 & $-$0.789 & 13.38 & 0.48  & 224   \\
        56  & 79.251 & 0.803  & 11.88 & 0.58  & 202   \\
        57  & 79.254 & 0.104  & $-$2.30 & 0.89  & 280   \\
        58  & 79.269 & 0.250  & 4.01  & 1.12  & 135   \\
        59  & 79.297 & $-$0.528 & 10.44 & 1.13  & 684   \\
        60  & 79.297 & 0.916  & 11.56 & 0.85  & 254   \\
        61  & 79.300 & 0.049  & 9.72  & 1.19  & 198   \\
        62  & 79.308 & $-$0.260 & $-$5.36 & 0.45  & 138   \\
    \hline
\end{tabular}
\end{table}

\setcounter{table}{0}

\begin{table}
    \centering
    \caption{Continued.}
    \begin{tabular}{cccccc}
    \hline
        ID   &
        \begin{tabular}[c]{@{}c@{}}$l$\\ {[} deg. {]}\end{tabular} & \begin{tabular}[c]{@{}c@{}}$b$\\ {[} deg. {]}\end{tabular} & \begin{tabular}[c]{@{}c@{}}$v_\mathrm{LSR}$\\ {[} \kms {]}\end{tabular} & \begin{tabular}[c]{@{}c@{}}$\sigma_v$\\ {[} \kms {]}\end{tabular} &
        \begin{tabular}[c]{@{}c@{}}$S_\mathrm{exact}$\\ {[} pix.$^2$ {]}\end{tabular}  \\
    \hline \hline
        63  & 79.337 & $-$1.020 & 6.67  & 0.94  & 124   \\
        64  & 79.364 & $-$0.472 & 0.93  & 0.78  & 868   \\
        65  & 79.400 & $-$0.636 & 4.68  & 1.48  & 5631  \\
        66  & 79.422 & $-$0.720 & 12.94 & 0.98  & 625   \\
        67  & 79.457 & 0.938  & $-$2.32 & 1.08  & 3937  \\
        68  & 79.476 & $-$0.041 & 7.79  & 1.30  & 473   \\
        69  & 79.504 & 0.970  & 11.83 & 0.97  & 536   \\
        70  & 79.507 & $-$0.979 & 6.93  & 0.81  & 116   \\
        71  & 79.645 & 0.346  & 5.73  & 0.59  & 237   \\
        72  & 79.661 & $-$0.721 & 10.74 & 0.64  & 165   \\
        73  & 79.681 & $-$0.360 & 3.39  & 0.71  & 423   \\
        74  & 79.710 & $-$0.884 & $-$2.03 & 0.57  & 215   \\
        75  & 79.759 & $-$0.636 & $-$1.28 & 0.23  & 147   \\
        76  & 79.787 & $-$0.482 & 0.11  & 0.71  & 149   \\
        77  & 79.813 & 1.010  & $-$7.50 & 0.72 & 421   \\
        78  & 79.836 & 0.890  & 10.76 & 0.63  & 447   \\
        79  & 79.852 & 0.834  & $-$4.06 & 0.76  & 436   \\
        80  & 79.971 & 1.179  & 12.69 & 0.50  & 477   \\
        81  & 80.018 & $-$0.403 & 0.11  & 1.15  & 545   \\
        82  & 80.044 & 0.571  & 11.22 & 0.59  & 426   \\
        83  & 80.099 & 0.617  & 3.78  & 1.24  & 1064  \\
        84  & 80.165 & $-$0.411 & $-$1.21 & 0.68  & 146   \\
        85  & 80.314 & 0.268  & $-$1.09 & 1.06  & 407   \\
        86  & 80.334 & 0.708  & $-$1.66 & 0.82  & 196   \\
        87  & 80.347 & 0.558  & $-$4.10 & 0.97  & 254   \\
        88  & 80.377 & 0.442  & 8.37  & 0.87  & 137   \\
        89  & 80.388 & 0.700  & 5.57  & 0.77  & 171   \\
        90  & 80.451 & 0.274  & 7.34  & 0.67  & 793   \\
        91  & 80.561 & 1.031  & $-$1.75 & 0.92  & 419   \\
        92  & 80.599 & 1.210  & 13.45 & 0.35  & 422   \\
        93  & 80.719 & 0.835  & 3.68  & 1.01  & 484   \\
        94  & 80.736 & $-$0.581 & $-$4.45 & 0.97  & 748   \\
        95  & 80.776 & 0.511  & $-$1.97 & 1.40  & 4231  \\
        96  & 80.801 & $-$0.337 & 5.48  & 1.74  & 13787 \\
        97  & 80.811 & 0.375  & 5.66  & 0.68  & 420   \\
        98  & 80.850 & 0.560  & 11.88 & 0.99  & 143   \\
        99  & 80.853 & $-$0.372 & 11.01 & 0.75  & 339   \\
        100 & 80.905 & $-$0.310 & $-$2.74 & 1.63  & 643   \\
        101 & 80.931 & 1.060  & 9.64  & 0.67  & 118   \\
        102 & 80.936 & 1.444  & 6.56  & 0.78  & 154   \\
        103 & 81.066 & 1.301  & 3.81  & 0.62  & 192   \\
        104 & 81.142 & 0.127  & 10.06 & 0.89  & 709   \\
        105 & 81.145 & 0.749  & 2.47  & 0.47  & 241   \\
        106 & 81.181 & 1.315  & $-$0.89 & 0.47  & 438   \\
        107 & 81.243 & 0.621  & 7.89  & 4.90  & 14139 \\
        108 & 81.246 & 1.437  & 4.95  & 0.47  & 620   \\
        109 & 81.361 & $-$0.029 & $-$5.00 & 1.49  & 3786  \\
        110 & 81.370 & 1.454  & $-$3.96 & 0.63  & 283   \\
        111 & 81.371 & 1.237  & 10.29 & 1.69  & 184   \\
        112 & 81.423 & $-$0.145 & 11.30 & 0.48  & 153   \\
        113 & 81.428 & 0.941  & $-$5.27 & 0.59  & 131   \\
        114 & 81.482 & $-$0.181 & 5.53  & 0.55  & 403   \\
        115 & 81.482 & 1.369  & $-$4.25 & 0.39  & 274   \\
        116 & 81.523 & 1.331  & 11.77 & 1.06  & 297   \\
        117 & 81.547 & $-$0.066 & $-$0.70 & 0.86  & 104   \\
        118 & 81.631 & $-$0.254 & 4.74  & 0.40  & 356   \\
        119 & 81.734 & 0.964  & 0.11  & 3.04  & 11051 \\
        120 & 81.771 & 0.344  & $-$5.89 & 0.43  & 313   \\
        121 & 81.839 & 0.255  & 8.65  & 1.66  & 12211 \\
        122 & 81.851 & 1.299  & 11.30 & 1.10  & 1566  \\
        123 & 81.908 & 0.082  & 14.24 & 0.93  & 865   \\
        124 & 82.061 & 0.924  & $-$3.96 & 0.57  & 91   \\
    \hline
\end{tabular}
\end{table}
\newpage

\bsp	
\label{lastpage}
\end{document}